%% file: main.tex
\documentclass[pra,twocolumn,amsmath,amssymb,superscriptaddress,floatfix,nofootinbib,aps]{revtex4-2}

\usepackage{graphicx, color, graphpap}      
\usepackage{amsmath}
\usepackage{enumerate}
\usepackage{amssymb}
\usepackage{amsthm}
\usepackage{float}
\usepackage[colorlinks=true, citecolor=blue, linkcolor=red]{hyperref}
\usepackage[T1]{fontenc}
\usepackage{bbm}
\usepackage{dsfont}
\usepackage[linesnumbered,ruled,vlined]{algorithm2e}
\SetKwInput{kwInit}{Init}
\usepackage{mathtools}

\usepackage{tikz}
\usepackage{graphicx}
\usetikzlibrary{positioning}
\usepackage{graphicx}
\usepackage{xcolor}
\usepackage[caption=false]{subfig}
\usepackage[ruled,vlined]{algorithm2e}
\usepackage{siunitx}
\usepackage{pifont} 
\usepackage{physics}
\usepackage{thmtools}
\usepackage{bytefield}
\usepackage{thm-restate}
\usepackage{quantikz}
\setcitestyle{numbers}
\usepackage{pgfplots}
\usetikzlibrary{calc}
\usepackage{listings}
\usepackage[most]{tcolorbox}
\makeatletter
\providecommand{\@afterenddocumenthook}{}
\makeatother

\newcommand{\Propaq}{Propaq}
\newcommand{\PauliProp}{\textsc{pauli-prop}}
\newcommand{\PauliPropagation}{\text{PauliPropagation.jl}}
\newcommand{\MajoranaPropagation}{\text{MajoranaPropagation.jl}}
\newcommand{\pyrauli}{\textsc{pyrauli}}
\newcommand{\MonoProp}{\textsc{monoprop}}

\lstdefinestyle{inlinecodestyle}{
  basicstyle=\ttfamily\small,
  columns=fullflexible,
  keepspaces=true,
  breaklines=false,
}
\newcommand{\inlinecode}{\lstinline[style=inlinecodestyle]}

\renewcommand{\ket}[1]{| #1 \rangle}
\renewcommand{\bra}[1]{\langle #1 |}

\renewcommand{\vec}[1]{\boldsymbol{#1}}  

\renewcommand{\eqref}[1]{(\ref{#1})}

\newtheoremstyle{example}{\topsep}{\topsep}%
{}
{}
{\bfseries}
{:}
{   }
{\thmname{#1}\thmnumber{ #2}}
\theoremstyle{example}

\theoremstyle{definition}

\newtheorem*{theorem*}{Theorem}

\def\orcid#1{\kern -0.4em\href{https://orcid.org/#1}{\includegraphics[keepaspectratio,width=0.7em]{orcid_logo.pdf}}}

\usepackage{lipsum}

\definecolor{codebg}{HTML}{FAFAFC}      
\definecolor{codeframe}{HTML}{D3D8E0}   
\definecolor{codetitlebg}{HTML}{E9EDF3} 
\definecolor{codekeyword}{HTML}{0B5394} 
\definecolor{codebuiltin}{HTML}{7A3E9D} 
\definecolor{codestring}{HTML}{9A6400}
\definecolor{codecomment}{HTML}{6B7280}
\definecolor{codelineno}{HTML}{A8AEB8}

\lstdefinestyle{propaq}{
    mathescape=true,
    escapeinside={(@}{@)},
    basicstyle=\ttfamily\scriptsize,
    keywordstyle=\bfseries\color{codekeyword},
    keywordstyle=[2]\color{codebuiltin},
    emphstyle=\color{codebuiltin},
    commentstyle=\itshape\color{codecomment},
    stringstyle=\color{codestring},
    numberstyle=\ttfamily\tiny\color{codelineno},
    numbers=left,
    numbersep=9pt,
    xleftmargin=2.1em,
    xrightmargin=0.2em,
    aboveskip=0pt,
    belowskip=0pt,
    breaklines=true,
    breakatwhitespace=true,
    breakindent=1.2em,
    postbreak=\mbox{\textcolor{codelineno}{$\hookrightarrow$}\space},
    columns=fullflexible,
    keepspaces=true,
    showspaces=false,
    showstringspaces=false,
    showtabs=false,
    tabsize=4,
    upquote=true,
    captionpos=b,
    emph={propaq,Propagator,Hamiltonian,PauliSum,PauliString,Observable,
          Truncation,NoisePolicy,FlushSchedule,Surrogate,propagate,evolve,
          truncate,flush,expectation},
}

\lstdefinelanguage{propaqpython}[]{Python}{
    morekeywords=[2]{self,None,True,False,len,range,print,list,dict,float,int,
                     str,zip,enumerate,sum,abs,super,np,__init__},
}

\tcbset{
    codecap/.style={%
        title={\textbf{Listing~\thetcbcounter.}\hspace{0.4em}#1}},
    language/.style={listing options={style=propaq, language=#1}},
    shell/.style={listing options={style=propaq, language=bash,
                                   numbers=none, xleftmargin=0.4em}},
}
\newtcblisting[auto counter, number within=section]{code}[1][]{%
    listing only,
    language=propaqpython,
    enhanced, breakable,
    colback=codebg,
    colframe=codeframe,
    colbacktitle=codetitlebg,
    coltitle=black,
    fonttitle=\footnotesize,
    boxrule=0.5pt,
    titlerule=0pt,
    arc=2pt, outer arc=2pt,
    boxsep=0pt,
    left=2pt, right=2pt, top=5pt, bottom=5pt,
    lefttitle=6pt, righttitle=6pt, toptitle=3pt, bottomtitle=3pt,
    title after break={\textbf{Listing~\thetcbcounter.}\hspace{0.4em}(continued)},
    #1
}

\long\def\ca#1\cb{} 
\begin{document}
\title{Propaq: A Python package for Heisenberg Propagation}

\author{Hrishikesh Belagali}
\affiliation{Department of Computational Mathematics, Science, and Engineering, Michigan State University, East Lansing, MI 48823, USA}
\affiliation{Center for Quantum Computing, Science, and Engineering, Michigan State University, East Lansing, MI 48823, USA}

\author{Ryan LaRose}
\thanks{Corresponding author:  \href{rmlarose@msu.edu}{rmlarose@msu.edu}}
\affiliation{Department of Computational Mathematics, Science, and Engineering, Michigan State University, East Lansing, MI 48823, USA}
\affiliation{Department of Electrical and Computer Engineering, Michigan State University, East Lansing, MI 48823, USA}
\affiliation{Department of Physics and Astronomy, Michigan State University, East Lansing, MI 48823, USA}
\affiliation{Center for Quantum Computing, Science, and Engineering, Michigan State University, East Lansing, MI 48823, USA}

\begin{abstract}
    We introduce Propaq, a performant and flexible Python package implementing Heisenberg propagation techniques for classically simulating quantum circuits. Propaq is compatible with major quantum software packages like Qiskit and Cirq, and matches or outperforms similar state-of-the-art packages for Heisenberg propagation. Beyond this, Propaq has multiple distinguishing features for simulation and research. These include Heisenberg propagation in any basis,  hybrid Schr\"{o}dinger-Heisenberg simulation, custom truncation and noisy simulation methods, and detailed logging information for analyzing simulation accuracy, runtime, and memory usage. We describe these features, how to use Propaq for research, and its backend implementation.
\end{abstract}

\maketitle


\section{Introduction}

Classically simulating quantum circuits is important practically to benchmark algorithms, error mitigation/correction techniques, and other protocols. Classical simulation is also important theoretically to refine the boundaries of quantum advantage~\cite{LaRose_2026} and ultimately to understand the power of quantum computation. In this vein, multiple famous theorems about quantum computation have been discovered through the lens of simulating circuits. Two prominent examples are the Gottesman-Knill theorem, which states that an important subclass of quantum operations known as Clifford operations are efficiently simulable~\cite{gottesmanHeisenbergRepresentationQuantum1998}, and the time evolving block decimation algorithm, which shows that weakly entangled one-dimensional circuits are efficiently simulable~\cite{vidal_efficient_2003}. Additional theorems have been proved in this context, especially recently for efficient classical simulation with noisy quantum circuits~\cite{gao_efficient_2018,Aharonov_Gao_Landau_Liu_Vazirani_2023,schuster2024,Nelson_Rajakumar_Hangleiter_Gullans_2026}.

Recently, the classical simulation technique underlying the Gottesman-Knill theorem has gained traction due to challenges to quantum advantage claims~\cite{kim_evidence_2023,Begusic_Gray_Chan_2024,Begusic_Chan_2025}. The simulation technique is commonly known as either sparse Pauli dynamics~\cite{Begusic_Gray_Chan_2024,Begusic_Chan_2025} or Pauli propagation~\cite{rudolph2025a}. Subsequently, Majorana propagation --- the same underlying technique but working in the Majorana basis instead of the Pauli basis --- was introduced~\cite{miller2025a}. Software packages for these techniques closely followed, with implementations in major quantum software packages like Qiskit and other standalone/plugin packages appearing. Software packages include the \PauliPropagation{}~\cite{rudolph2025a} and \MajoranaPropagation{}~\cite{danna2025majorana} Julia packages, the Paulistrings.jl Julia package~\cite{loizeau_quantum_2025}, the \PauliProp{} Qiskit add-on~\cite{pauli-prop}, the \pyrauli{} package~\cite{zeFresk_2026} which is implemented in C with Python bindings, and most recently the \MonoProp~\cite{miller2025a,monoprop-github,monoprep} package which supports both Pauli and Majorana propagation.

\begin{figure*}
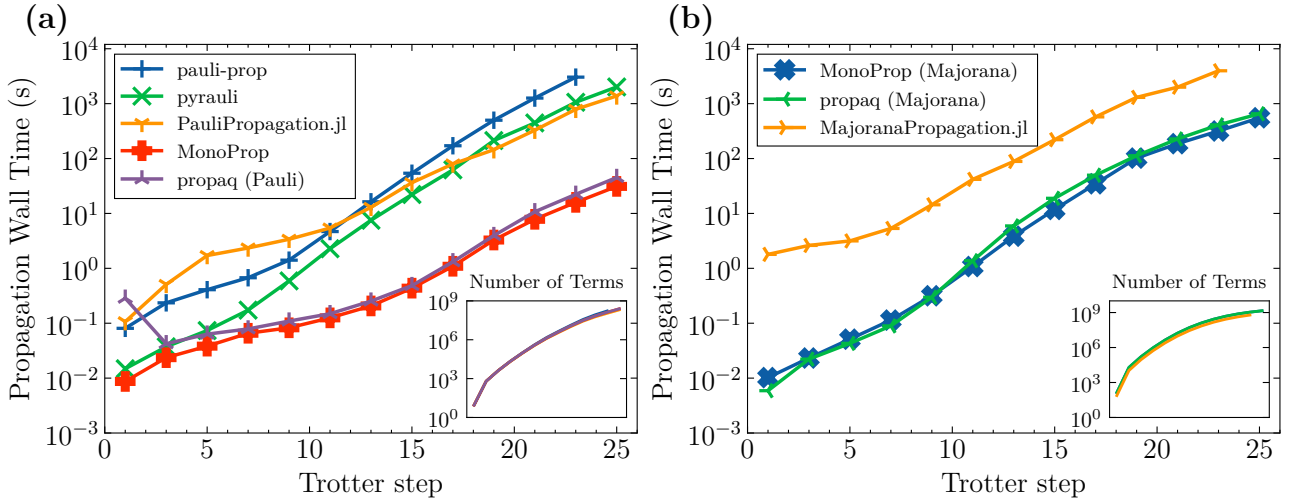

    \begingroup
    \providecommand{\mathdefault}[1]{#1}
    \catcode`\_=12
    \newcommand{\runtimepanel}[2]{%
        \begin{tikzpicture}
            \node[inner sep=0pt] (panelbox) {\resizebox{0.5\textwidth}{!}{\input{#2}}};
            \node[anchor=south west,font=\bfseries\large,inner sep=1pt] at ($(panelbox.north west)+(0.015\textwidth,-0.008\textwidth)$) {#1};
        \end{tikzpicture}%
    }
        \begin{tikzpicture}
            \node[inner sep=0pt] (panelbox) {\resizebox{0.5\textwidth}{!}{\input{figures/ising_runtime.pgf}}};
            \node[anchor=south west,font=\bfseries\large,inner sep=1pt] at ($(panelbox.north west)+(0.015\textwidth,-0.008\textwidth)$) {(a)};
        \end{tikzpicture}%
        \begin{tikzpicture}
            \node[inner sep=0pt] (panelbox) {\resizebox{0.5\textwidth}{!}{\input{figures/hubbard_runtime.pgf}}};
            \node[anchor=south west,font=\bfseries\large,inner sep=1pt] at ($(panelbox.north west)+(0.015\textwidth,-0.008\textwidth)$) {(b)};
        \end{tikzpicture}%
    
    \caption{Heisenberg propagation wall-clock runtime and number of terms (inset) for various packages. Here, we benchmark computing expectation values of a $Z$ observable for 
    (a) Trotterized time evolution of a 2D TFIM on a $6\times 6$ square lattice ($J = 1.0$, $h = 0.5$, $\Delta t = 0.1$) and
    (b) Trotterized time evolution of a 2D Hubbard model on a $3 \times 3$ square lattice ($t = 1.0$, $U = 4.0$, $\Delta t = 0.1$). All backends use a coefficient cutoff of $10^{-6}$ and 64 threads, except for the single-threaded \PauliProp{} package. As shown, \Propaq{} performs similarly to \PauliPropagation{} and \MonoProp{} (in the Pauli basis), and performs similarly to \MonoProp{} (in the Majorana basis). We remark that \Propaq{} adopts several low-level design elements from \MonoProp{}~\cite{monoprop-github,monoprep} to achieve this performance, described in Sec.~\ref{sec:design}.}
    \label{fig:ising-trotter-runtime}
    \endgroup
\end{figure*}

In this work, we introduce a Python package for Heisenberg propagation that we call \Propaq{}. \Propaq{} is implemented in Rust for performance and has Python front-ends for usability with major quantum software packages. Notably, direct support for Qiskit circuits is provided. We show in numerical benchmarks that \Propaq{} is among the fastest Heisenberg simulators available. Further,  \Propaq{} has several features that distinguish it from current state of the art. Notably, \Propaq{} supports Heisenberg propagation in any basis. The Pauli and Majorana bases are built-in to the library and natively supported, and any other user-defined bases are also supported. \Propaq{} also supports hybrid Schr\"{o}dinger-Heisenberg simulation. Here, instead of back-propagating observables fully through the circuit, we allow forward evolution to a bipartation. In certain cases, this can provide a useful space/time tradeoff and present advantages in simulation, as was explored recently in quantum-classical hybrid simulation~\cite{fuller_improved_2025}. Additionally, \Propaq{} contains logging tools to easily analyze simulation accuracy, providing details on the total number of terms and total $L_1$ coefficient norm discarded, among other metrics. \Propaq{} also supports custom noise channels and natively supports noisy simulation with any propagation method. For these reasons, we expect \Propaq{} to be a useful package for quantum information researchers and software developers, enabling both fast simulation and future research in (hybrid Schr\"{o}dinger-)Heisenberg simulation.

To these ends, the rest of the paper is organized as follows. We first provide more details on both the features and performance of \Propaq{} mentioned above, showing benchmarks with other packages and highlighting \Propaq{}'s capabilities, in Sec.~\ref{sec:features}. We then provide background on Heisenberg propagation in Sec.~\ref{sec:background}. Section~\ref{sec:api} shows how to get started with \Propaq{} and how to use the library through various examples. In Sec.~\ref{sec:design}, we discuss the design and implementation of \Propaq{}, providing backend implementation and low-level software details, including designs adopted from \MonoProp{}~\cite{monoprop-github,monoprep} that are crucial for performance. Finally, we provide more performance benchmarks for \Propaq{} in Sec.~\ref{sec:performance}.

\section{Features and Performance} \label{sec:features}

\subsection{Benchmarks with other packages}

In Fig.~\ref{fig:ising-trotter-runtime}, we evaluate the performance of \Propaq{} against other packages for Heisenberg propagation, including \PauliProp{}, \pyrauli{}, \PauliPropagation{}, \MajoranaPropagation{}, and \MonoProp{}. 
In this benchmark, we show the wall-clock runtime for computing the expectation value of a single $Z$ observable after Trotterized time evolution of a $6 \times 6$ transverse-field Ising model (Fig.~\ref{fig:ising-trotter-runtime}(a))
and a $3 \times 3$ Hubbard model (Fig.~\ref{fig:ising-trotter-runtime}(b)). For all benchmarks, we utilize a coefficient cutoff of $10^{-6}$ and run on 64 AMD EPYC 7H12 cores. When applicable, we use both Pauli and Majorana propagation. Results are shown in Fig.~\ref{fig:ising-trotter-runtime}. As can be seen \Propaq{} performs similarly to \MonoProp{} and \PauliPropagation{}. To achieve this performance, \Propaq{} adopts several low-level design elements for term storage and threading that are critical for performance from \MonoProp{}~\cite{monoprop-github,monoprep}, described in Sec.~\ref{sec:design}. While packages have different strengths and performance is somewhat benchmark dependent, we have found this behavior to be generally true for all packages. It is worth noting that \PauliProp{} has a single-threaded implementation while other packages are multi-threaded. In this benchmark, all packages maintain a
similar number of terms in the back-propagated operator, and the multi-threaded packages consequently have similar peak memory usage. Memory usage is shown in more detail in Fig.~\ref{fig:ising-trotter-memory} in Sec.~\ref{sec:performance}, which contains additional performance benchmarks.

\subsection{Arbitrary basis backpropagation} \label{subsec:arbitrary-basis}

\begin{figure}
    \centering
    \begingroup
    \providecommand{\mathdefault}[1]{#1}
    \resizebox{\columnwidth}{!}{\input{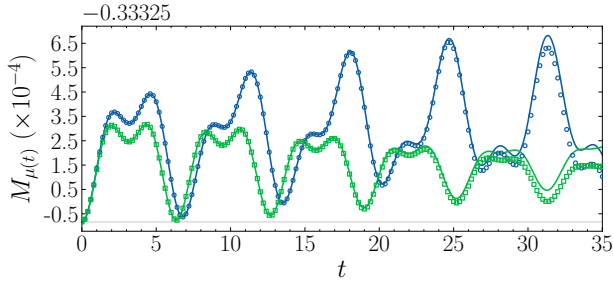}}
    \endgroup
    \caption{An illustrative example of \Propaq{}'s ability to perform Heisenberg evolution in any basis, demonstrated via time evolution of the three-state (qutrit) Potts model~\cite{krasznai2026}. We subclass \Propaq{}'s core propagation engine to do Heisenberg evolution with Weyl-Heisenberg operators, a higher-dimensional generalization of qubit Pauli operators. Here we compute the magnetizations $M_2(t)$ (blue) and $M_3(t)$ (green) for a quench of the three-state Potts spin chain in the positive oblique regime ($h_2=0.1$, $n=7$) using exact diagonalization (solid) and Weyl-Heisenberg propagation (markers) in \Propaq{}. After creating a subclass to enable propagation in a new basis, all features of \Propaq{} can be used. To illustrate this, we perform backpropagation with a weight cutoff of $w = 4$. As can be seen, results are in excellent agreement with exact diagonalization up to $t \le 25$ despite the weight cutoff, then start to accumulate error after. When computing these observables, the system is initialized in the fully polarized state.}
    \label{fig:potts-dynamics}
\end{figure}
Although Heisenberg propagation is generally carried out in the Pauli or Majorana basis, it applies to any operator basis. \Propaq{} supports this natively in its design. To implement backpropagation in any basis, the following methods need to be implemented:
\begin{enumerate}
    \item Products of basis elements,

    \item If a term branches under a circuit operation,

    \item The operator weight (for truncation support),

    \item Diagonal expectation with a basis state,

    \item Express gates in terms of basis generators.
    
    \item Hashing, equality and byte representations for term I/O
\end{enumerate}
These are defined by subclassing \Propaq{}'s abstract classes for propagation and implementing the required abstract methods. Example code is provided in Sec.~\ref{sec:api}. After this, observable backpropagation can be used as with this custom basis in the same way as with the built-in bases (Pauli and Majorana).

As an example, we use \Propaq{} to compute magnetizations in the Potts model in Fig.~\ref{fig:potts-dynamics}. Here, we use backpropagation with Weyl-Heisenberg operators, a generalization of Pauli operators to qudits. Creating a subclass for Weyl-Heisenberg propagation allows us to use \Propaq{} for simulating circuits with all features, notably term truncation. In Fig.~\ref{fig:potts-dynamics}, the magnetizations of the seven-site, three-state Potts spin chain are computed with a weight cutoff of $w = 4$. As shown, the computed expectation values are in excellent agreement with exact diagonalization for short times, up to $t \le 25$. Afterwards, the expectation values accumulate error due to the $w=4$ truncation. The complete code in \Propaq{} to implement this example is provided in Appendix~\ref{sec:weyl-propagation}. To the best of our knowledge, \Propaq{} is the only library to support backpropagation in any basis, and this example would not be possible with other packages.

\subsection{Hybrid Schr\"{o}dinger-Heisenberg simulation}

\begin{figure}
    \centering
    \includegraphics[width=\columnwidth]{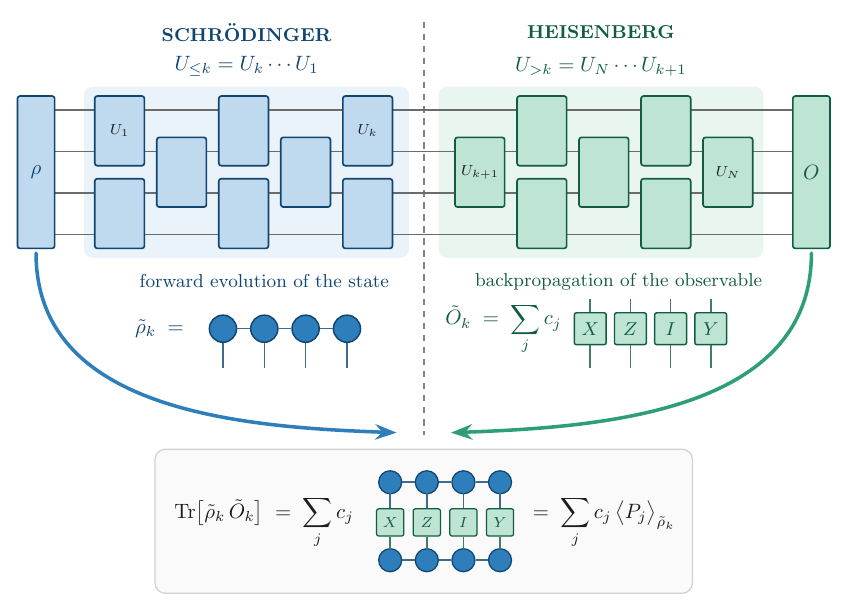}
    \caption{Graphical illustration of the Schr\"{o}dinger-Heisenberg simulation method~\eqref{eqn:hybrid-algorithm}. The circuit is partitioned at index $k$ (dashed line). Operations before the partition are applied to the initial state $\rho$ via Schr\"{o}dinger simulation, for example a matrix product state (MPS). Operations after the partition are applied to the observable $O$ via Heisenberg simulation, for example in the Pauli basis as illustrated. Finally, the expectation value is computed as a series of overlaps between the forward-evolved state and the backpropagated observable. In \Propaq{}, forward MPS is currently implemented, but custom methods are support --- all that needs to be defined is how to compute overlaps between the forward-evolved state and the backpropagated observable basis elements.}
    \label{fig:schrodinger-heisenberg-schematic}
\end{figure}

In addition to traditional Heisenberg propagation, \Propaq{} supports a hybrid Schr\"{o}dinger-Heisenberg simulation algorithm. In this approach, a circuit is divided into two parts: the first part is simulated in the Schr\"{o}dinger picture, and the latter is simulated in the Heisenberg picture. In other words, for circuit $U = U_N \cdots U_1$, this method evaluates expectation values via
\begin{equation} \label{eqn:hybrid-algorithm}
    \Tr \left[ U \rho U^\dagger O  \right]
    = \Tr \left[ \tilde{\rho}_k \tilde{O}_k \right],
\end{equation}
where
\begin{equation}
    \tilde{\rho}_k :=U_k \cdots U_1 \, \rho \, U_1^\dagger \cdots U_k^\dagger
\end{equation}
and
\begin{equation}
    \tilde{O}_k := U_{k+1}^\dagger \cdots U_N^\dagger \, O \, U_N \cdots U_{k+1} .
\end{equation}
This is shown schematically in Fig.~\ref{fig:schrodinger-heisenberg-schematic}. The expectation value~\eqref{eqn:hybrid-algorithm} is computed via a series of overlap computations between $\tilde{\rho}_k$ and the basis-expanded $\tilde{O}_k$. Currently, \Propaq{} supports matrix product states for the forward Schr\"{o}dinger evolution, but in principle any method is supported. Indeed, the user only needs to be define the method for forward simulation (which is commonly done via another package, e.g. Stim for Clifford simulation~\cite{gidney_stim_2021}) and the overlap calculation for the particular basis. We expect this hybrid algorithm to present a useful space-time tradeoff for particular problems and/or compute platforms, and to our knowledge \Propaq{} is the only Heisenberg simulation package to support this. We note that this hybrid algorithm is also compatible with forward evolution on a quantum computer as in~\cite{fuller_improved_2025}. Code for using this hybrid algorithm method is shown in detail in Sec.~\ref{sec:api}.

\subsection{Truncation, logging, and accuracy}

\begin{figure}
    \centering
    \includegraphics[width=\linewidth]{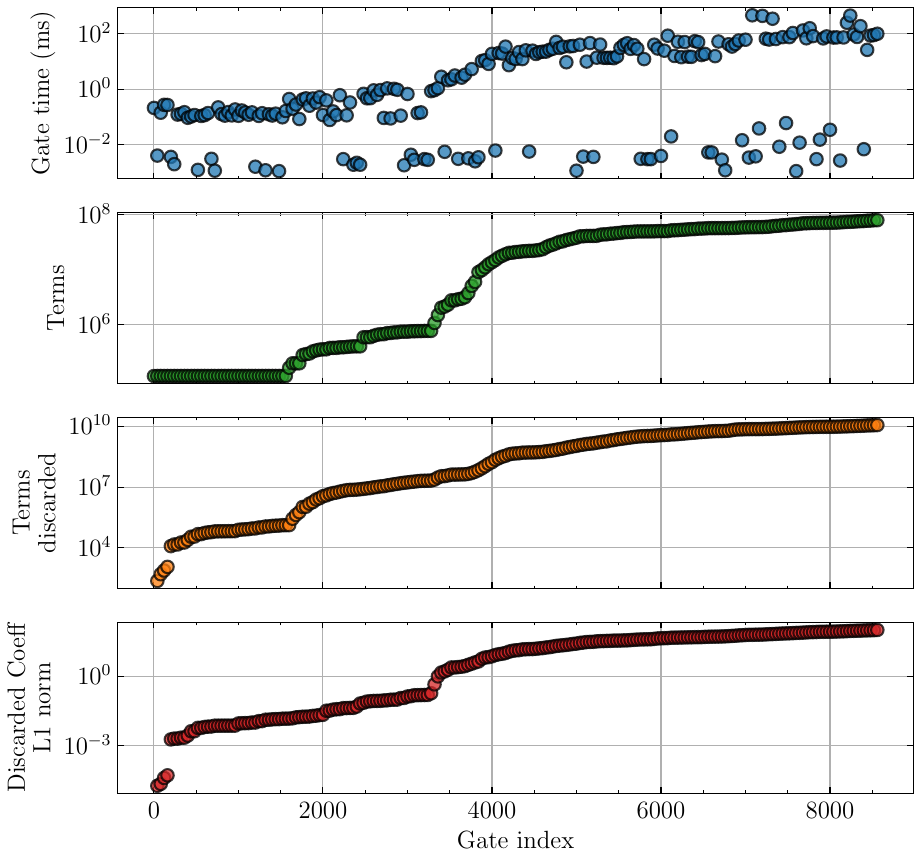}
    \caption{Illustration of some logging/analysis features in \Propaq{}. From top to bottom, this plot shows the time to apply each gate in milliseconds, the cumulative number of terms in the simulation, the cumulative number of terms discarded, and the cumulative $L_1$ norm of discarded coefficients. The last metric provides an upper bound on observable accuracy. \Propaq{} contains additional logging/analysis metrics not shown here, for example truncation events, total elapsed time, and maximum discarded coefficient $L_1$ norm. The data shown is for the simulation of a two-layer unitary cluster Jastrow ansatz~\cite{motta2023} with Majorana propagation and a coefficient cutoff of $10^{-6}$.}
    \label{fig:logging}
\end{figure}

A cornerstone of Heisenberg simulators is truncation, used as a heuristic to circumvent the general exponential complexity of the simulation. While this is commonly employed, it is not well understood how to choose the best truncation strategy for a given problem, or what the observable accuracy will be. Indeed, several methods have been introduced and explored for truncation including Clifford perturbation theory~\cite{begusic_simulating_2023}, state-dependent projections~\cite{perez_state-adapted_2026}, and stochastic methods~\cite{zhang_stochastic_2026}. To address these issues, \Propaq{} enables the use of both commonly used truncation strategies --- coefficient and weight truncation --- plus custom truncation strategies. 

Truncation is handled in \Propaq{} via a sequence of truncation strategies, as below.

\begin{code}[language=python, codecap={Truncation policies in \Propaq{}.}, label={lst:truncation-policies-features}]
from propaq.truncation import WeightTruncator, CoefficientTruncator, TermBudget

truncation = [
    CoefficientTruncator(1e-6),
    WeightTruncator(5),
    TermBudget(min_terms=1_000)
]
\end{code}

\noindent This truncation strategy says to discard terms with coefficient magnitude smaller than $10^{-6}$, discard terms with weight larger than five, and never drop below one thousand terms. Any combination of these policies can be used in simulation, and custom truncation policies can be defined. Truncation policies are incorporated into propagators at instantiation.

\begin{code}[language=python, codecap={Syntax for propagation with truncation.}, label={lst:propagation-with-truncation}]
from propaq import PauliPropagator

prop = PauliPropagator(truncation=truncation)
\end{code}
\noindent
The use of propagators to simulate circuits is done by calling the \inlinecode{propagate} or \inlinecode{expectation_value} methods, shown in Sec.~\ref{sec:api}. We remark that custom propagators in other bases (Sec.~\ref{subsec:arbitrary-basis}) support truncation in the same manner.

On top of truncation, \Propaq{} contains detailed logging tools to analyze runtime and memory complexity as well as simulation accuracy. First, progress through a simulation can be tracked during runtime with a progress bar.

\begin{code}[language=python, codecap={\Propaq{} implements a progress bar to show simulation progress and estimate total runtime.}, label={lst:propagation-with-progress-bar}]
from propaq import PauliPropagator

propagator = PauliPropagator(progress_bar=True)
result = propagator.expectation_value(...)
# Propagating: 100% 44/44 [00:00<00:00, 273.21gate/s, terms=50]
\end{code}
\noindent As shown, progress bars show the percentage of gates applied in the circuit, gate application time, and total number of terms in the simulation. The returned \inlinecode{result} contains the number of terms in backpropagated observable after every gate applied in simulation, which is useful for analyzing term growth and simulation complexity. 

Even more detailed analyses of the simulation can be enabled with \Propaq{}'s logging tools. One can log simulation statistics by creating a propagator with a logger object as below.
\begin{code}[language=python, codecap={Logging simulation statistics in \Propaq{}.}, label={lst:propagation-with-logging}]
from propaq import CoefficientTruncator, Logger, MajoranaPropagator

logger = Logger("log.jsonl", log_every=1)
prop = MajoranaPropagator(
    truncation=CoefficientTruncator(1e-6),
    progress_bar=True,
    logger=logger,
)
\end{code}

\noindent This provides the propagator with logging tools to track timing of various stages of the propagation, as well as truncation statistics, including the $L_1$ norm of the discarded terms, number of terms discarded, largest magnitude of discarded terms, etc. An example of this functionality is shown in Fig.~\ref{fig:logging} for simulating a chain of twenty Hydrogen atoms with Pauli propagation and a cutoff of $10^{-6}$. Here, we see that, even though nearly $10^8$ terms are discarded due to truncation, the discarded coefficient $L_1$ norm is more mild, between $10^0$ and $10^1$, providing a bound on observable accuracy. 
In addition to providing insight to simulation performance and accuracy, these tools are provided to aid in fine-tuning truncation parameters, and to identify bottlenecks in custom propagation engines.

Finally, we note that \Propaq{} provides API methods to
save backpropagated operators to disk in a compressed format. Operators can be loaded from disk into memory all at once,
or in a streaming fashion, where terms are loaded one at a time. This allows for distributed computation of expectation values across
multiple devices, as well as the ability to perform checkpoint and resume simulation. Code for saving and (lazily) loading operators so is shown in Sec.~\ref{sec:api}.

\subsection{Noisy simulation}

In addition to truncation, another cornerstone of Heisenberg propagation techniques is noisy simulation. The reason for this is that noise can make simulation easier~\cite{schuster2024}. In particular, depolarizing noise has the effect of damping coefficients, as reviewed in Sec.~\ref{sec:background}, by causing the coefficients of some terms to go to zero (or below the truncation threshold), reducing the total number of terms in the backpropagated observable and thus reducing the complexity of the simulation.

\Propaq{} contains native support for noisy simulation with implementations of depolarizing noise.

\begin{code}[language=python, codecap={Native noise models in \Propaq{}.}, label={lst:native-noise}]
import propaq
from propaq import UniformNoiseModel

noise_model = UniformNoiseModel(damping=0.001)

# Pass into a Pauli/Majorana propagator
prop = propaq.PauliPropagator(noise=noise_model)
\end{code}

\noindent The \inlinecode{UniformNoiseModel} applies single-qubit depolarizing noise to all qubits after every layer of operations.
\Propaq{} also defines \inlinecode{GateNoiseModel} which allows the user to specify the noise channel to apply.  Beyond this, custom noise channels are supported and can be used with all \Propaq{} propagation features. An example of defining a custom channel for use in noisy simulation is Sec.~\ref{sec:api}.

\subsection{Extrapolation Methods}

\begin{figure}
    \centering
    \begingroup
    \providecommand{\mathdefault}[1]{#1}
    \catcode`\_=12
    \resizebox{\columnwidth}{!}{\input{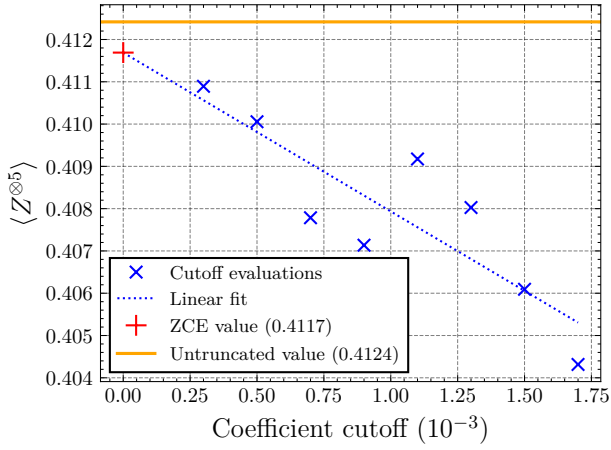}}
    \endgroup
    \caption{Zero-coefficient-cutoff extrapolation for a five-qubit periodic TFIM circuit evolved for 4 Trotter steps. A linear extrapolation is applied to estimate the untruncated expectation value of $Z^{\otimes 5}$. As can be seen, the extrapolated value is closer to the exact result than the truncated values.}
    \label{fig:zce}
\end{figure}

Both truncation and noisy simulation allow for the implementation of extrapolation methods. In particular, \Propaq{} supports zero-cutoff extrapolation (ZCE) in which the simulation is performed at increasing truncation levels, then a model is fit to the data to predict the zero-cutoff limit behavior of simulation. 
In Fig.~\ref{fig:zce}, we show that a truncated expectation value can be extrapolated to the zero-cutoff limit for a five-qubit TFIM circuit, improving the simulation accuracy. The API methods for implementing this extrapolation technique are described in Sec.~\ref{sec:api}. 

Since noisy simulation generally makes Heisenberg propagation easier, as mentioned above, it is also natural to consider zero-noise extrapolation~\cite{temme2017, giurgica-tiron_digital_2020} in this context. Here, the simulation is performed at increasing noise levels (with decreasing simulation cost), then a model is fit to the data and extrapolated to the zero-noise limit. The API methods for implementing this extrapolation technique are also described in Sec.~\ref{sec:api}.

\subsection{Optimizations}\label{sec:optimizations}

\begin{figure}
    \centering
    \begingroup
    \providecommand{\mathdefault}[1]{#1}
    \catcode`\_=12
    \resizebox{\columnwidth}{!}{\input{figures/surrogate.pgf}}
    \endgroup
    \caption{Optimization wall time for depth-two QAOA max-cut ansatz on a 12-vertex, 3-regular graph. 1000 iterations of direct numerical propagation is compared to surrogate model build time (orange) and evaluation time (green). While symbolically simulating the circuit takes 1.34 seconds, after this the total evaluation time for all optimization iterations is 1.18 seconds, providing a significant speedup over optimizing with direct numerical evaluation (12.2 seconds).}
    \label{fig:surrogate-optimization}
\end{figure}
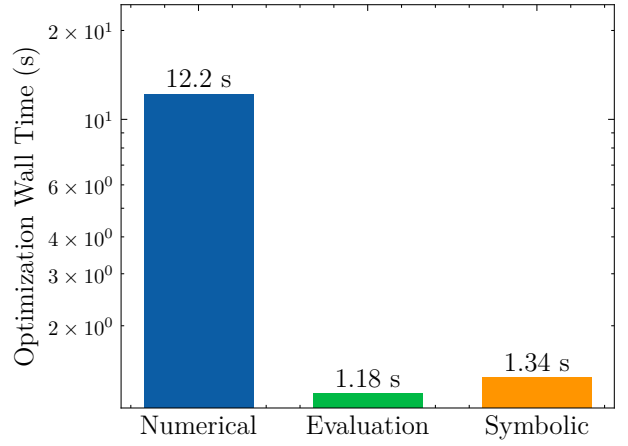

\Propaq{} contains multiple optimizations for performance, both at the software and algorithmic/theory level. While the low-level backend design is described in Sec.~\ref{sec:design}, we highlight two algorithmic optimizations here. First, tuning (variationally optimizing) gate parameters is a common algorithmic primitive in the variational quantum eigensolver~\cite{peruzzo_variational_2014}, quantum approximate optimization algorithm~\cite{farhi_quantum_2014}, and other variational quantum algorithms~\cite{khatri_quantum-assisted_2019,larose_variational_2019,bravo-prieto_variational_2023,cerezo_variational_2021}. In this setting, circuits need to be simulated many times as a subroutine to optimize the objective function. To mitigate this problem, \Propaq{} supports symbolic/surrogate propagation. In this method, the objective function is computed once with symbolic angles, then can be efficiently evaluated for other parameters. While the overhead of the first simulation can be large depending on the problem, this technique can present significant utility when the number of optimization steps is large and renders serial simulation at separate parameters intractable. An example of this is shown in Fig.~\ref{fig:surrogate-optimization}. Here, we fix 1000 iterations for a depth-two QAOA max-cut ansatz on twelve qubits. The total time for optimization numerically (non-symbolic simulation) is 12.2 seconds. While symbolically simulating the circuit the first time takes 1.34 seconds, after this the total evaluation time for all optimization iterations is 1.18 seconds, providing a significant speedup over direct numerical evaluation.

\begin{figure}
    \centering
    \begingroup
    \providecommand{\mathdefault}[1]{#1}
    \catcode`\_=12
    \resizebox{\columnwidth}{!}{\input{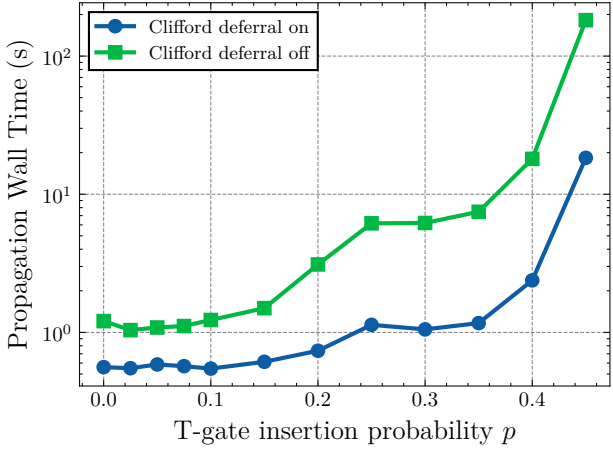}}
    \endgroup
    \caption{Runtime comparison of $\langle Z \rangle$ calculation with and without Clifford deferral on a 64-qubit, 80 layer Clifford+$T$ circuit.
    Each layer of the circuit consists of a random single-qubit Clifford gate on each qubit, followed by a perfect-matching layer of 
    CNOT gates. A $T$ gate is applied to a random qubit with probability $p$. As shown, \Propaq{}'s Clifford deferral optimization (Sec.~\ref{sec:optimizations}) can lead to a large decrease in runtime, up to around 10x, especially for circuits with high $T$ gate density.}
    \label{fig:clifford-deferral}
\end{figure}

Additionally, \Propaq{} implements another algorithmic optimization for simulating circuits that we call Clifford deferral.
In general, for each gate application in the circuit we will have to evaluate two terms, 
$\cos(\theta)$ and $\sin(\theta)$, where the angle $\theta$ is determined by the term and gate (see~\eqref{eqn:gate-update}). This potentially requires allocating and re-merging a large number of terms. For Clifford gates, 
we can store the adjoint action of the gate in a \emph{stabilizer tableau}~\cite{aaronsonImprovedSimulationStabilizer2004}, and defer the action of a sequence of Clifford gates. Once a 
non-Clifford gate is encountered, it is pushed through the tableau and applied to all terms in parallel, providing efficiency gains.
To illustrate this, we perform a timing study, comparing the performance of \Propaq{} with and without Clifford deferral on random Clifford+T circuits. The results, shown in Fig.~\ref{fig:clifford-deferral}, demonstrate that deferring Clifford gates can confer significant performance gains for circuits with high densities of Clifford gates.

\section{Background} \label{sec:background}
\subsection{Heisenberg Propagation}
We briefly review the theory of Heisenberg propagation~\cite{gottesmanHeisenbergRepresentationQuantum1998,miller2025a,rudolph2025a,rudolph2026}.
Consider a quantum circuit
\begin{equation}
    \mathcal{C} \coloneq U_L \ldots U_2 U_1
\end{equation}
consisting of $L$ gates, an observable $\mathcal{O}$, and a quantum state $\ket{\Psi_0}$.
The expectation value of the observable with respect to the state after the circuit is applied is given by 
\begin{equation} \label{eqn:expectation-value-schrodinger}
\mathbb{E}[\mathcal{O}] := \bra{\Psi_0}{\mathcal{C}^\dagger \mathcal{O} \mathcal{C}}\ket{\Psi_0} = \Tr\left[ \mathcal{O} \left( \mathcal{C}\rho\mathcal{C}^\dagger \right) \right]
\end{equation}
with
\begin{equation}
    \rho \coloneq \ket{\Psi_0}\bra{\Psi_0} .
\end{equation}
This is the Schr\"odinger picture of quantum mechanics in which the state evolves in time and the observable is fixed. 
Through the cyclicity of the trace, we can equivalently write this as 
\begin{equation} \label{eqn:expectation-value-heisenberg}
    \mathbb{E}[\mathcal{O}] = \Tr \left[  \left( \mathcal{C}^\dagger \mathcal{O} \mathcal{C} \right) \rho \right]
\end{equation} 
to recover the Heisenberg picture, where the observable is \textit{back-propagated} through the circuit and the state remains fixed. 
By representing the observable in a basis of operators $\{\sigma_i\}$, we can write 
$$\mathcal{O} = \sum_i c_i \sigma_i$$
where $c_i$ are the coefficients of the observable in the operator basis. 
The action of the circuit is precisely the adjoint action of each layer on the basis elements, which can be computed via the commutation relations of the operator basis. 
After parameterizing each gate as 
\begin{equation}
    U_i = \exp{-i \theta \sigma_i / 2}  ,
\end{equation}
the action of the gate on a basis element $\sigma_j$ is given by
\begin{equation} \label{eqn:gate-update}
    \sigma_j \mapsto \begin{cases}
        \sigma_j & [\sigma_i, \sigma_j] = 0 \\
        \cos(\theta) \sigma_j + i \sin(\theta) \sigma_i'& [\sigma_i, \sigma_j] \neq 0
    \end{cases}
\end{equation}
 where $\sigma_i'$ is a new basis element generated by the commutator $[\sigma_i, \sigma_j]$. Because of this, the number of terms in the back-propagated observable will generally grow exponentially with the number of gates in the circuit, and it is often necessary to truncate the back-propagated observable to a manageable number of terms. 
 This is generally done by discarding terms with small coefficients or high weight.

\subsection{Noisy Heisenberg Propagation}

Heisenberg propagation with noise can lessen simulation complexity~\cite{schuster2024} and thus presents a common application. Here, we are interested in the computation of~\eqref{eqn:expectation-value-heisenberg} with noise applied to the state. That is, our goal is to estimate
\begin{equation}
    \mathbb{E} [\mathcal{O}] := \text{Tr} [ \mathcal{E} (U \rho U^\dagger) \mathcal{O} ] 
\end{equation}
where $\mathcal{E}$ is a quantum channel (completely positive, trace preserving map). Naturally, there are many such maps $\mathcal{E}$ defined by their action and when/where they act during the quantum computation. Most generally, a quantum channel is defined by its Kraus operators $\{ K_i \}_{i = 1}^{m}$ via
\begin{equation} \label{eqn:quantum-channel-kraus-operators}
    \mathcal{E} (\rho) = \sum_i K_i \rho K_i ^\dagger ,
\end{equation}
with the only restriction on $K_i$ being the completeness relation
\begin{equation}
    \sum_i K_i^\dagger K_i = I .
\end{equation}
Two typical models that have been studied are single-qubit depolarizing noise
\begin{equation} \label{eqn:depolarizing-channel}
    \mathcal{E}_p (\rho) := (1 - p) \rho + \frac{p}{3} \left( X\rho X + Y \rho Y + Z \rho Z \right)
\end{equation}
where $0 \le p \le 1$ and  $X$, $Y$, and $Z$ are the Pauli matrices applied (1) after every parallel layer of gates in the circuit, or (2) after every gate in the circuit. These  have been studied in~\cite{schuster2024} because they model quantum hardware relatively well but are still tractable to work with analytically. We remark that \Propaq{} implements (1) as \inlinecode{UniformNoiseModel}, as shown in Sec.~\ref{sec:api}.
Consider a single-qubit state $\rho$ acted on by a depolarizing channel~\eqref{eqn:quantum-channel-kraus-operators}. Our goal is to evaluate
\begin{equation}
    \text{Tr} [ \mathcal{E}_p (\rho) \mathcal{O} ] .
\end{equation}
Using the Kraus operator representation~\eqref{eqn:quantum-channel-kraus-operators} and simple properties of the trace, we may write
\begin{align}
    \text{Tr} [ \mathcal{E} (\rho) \mathcal{O} ] &= \text{Tr} \left[ \left( \sum_i K_i \rho K_i^\dagger \right) \mathcal{O} \right] \\
    &= \sum_i \text{Tr} \left[  K_i \rho K_i^\dagger \mathcal{O} \right] \\
    &= \sum_i \text{Tr} \left[  \rho K_i^\dagger \mathcal{O} K_i \right] \\
    &= \text{Tr} \left[  \rho \left( \sum_i K_i^\dagger \mathcal{O} K_i \right) \right] \\
    &=: \text{Tr} [ \rho \mathcal{E}^\dagger (\mathcal{O})  ] .
\end{align}
In the last step, we defined the adjoint (dual) channel $\mathcal{E}^\dagger$ with Kraus operators $\{ K_i^\dagger \}$. Certain channels, notably the single-qubit depolarizing channel~\eqref{eqn:depolarizing-channel}, are self-adjoint, meaning that $K_i^\dagger = K_i$ for all $i$ and so $\mathcal{E}^\dagger = \mathcal{E}$. 

In Pauli propagation, 
\begin{equation}
    \mathcal{O} = \sum_i c_i P_i 
\end{equation}
where $P_i$ are $n$-qubit Pauli operators of the form
\begin{equation} \label{eqn:pauli}
    P = s X_1^{x_1} X_2^{x_2} \cdots X_n^{x_n} Z_1^{z_1} Z_2^{z_2} \cdots Z_n^{z_n} .
\end{equation} 
Here, $s \in \mathbb{C}$ is a phase, $X$ and $Z$ are the single-qubit Pauli operators defined by
\begin{align}
    X &= \left[ \begin{matrix}
        0 & 1 \\
        1 & 0
    \end{matrix} \right] \\
    Z &= \left[ \begin{matrix}
        1 & 0 \\
        0 & -1
    \end{matrix} \right] ,
\end{align}
and $x_i, z_i \in \{0, 1\}$. Note that $Y := i XZ$ can be generated up to phase from the product of $X$ and $Z$, and that $X^2 = Y^2 = Z^2 = I$. Since $\mathcal{P} := \{I, X, Y, Z\}$ is a basis for any single-qubit operator, $\mathcal{P}_n := \mathcal{P}^{\otimes n}$ is a basis for any $n$-qubit operator. 

Consider a single-qubit Pauli $P$ under a single-qubit depolarizing channel $\mathcal{E}_p$. By definition, we have
\begin{equation}
    \mathcal{E}_p (P) = (1 - p) P + \frac{p}{3} (XPX + YPY + ZPZ).
\end{equation}
For any non-identity Pauli $P \in \{X, Y, Z\}$, it is easy to verify that
\begin{equation}
    XPX + YPY + ZPZ = -P .
\end{equation}
Thus, the action of the depolarizing channel on a non-identity Pauli can be written
\begin{equation}
    \mathcal{E}_p(P) = (1 - p) P + \frac{p}{3} (-P) = \left( 1 - \frac{4}{3} p \right) P. 
\end{equation}
For $P = I$, we have that
\begin{equation}
    \mathcal{E}_p(I) = I .
\end{equation}
Therefore, when propagated through a channel, non-identity Paulis pick up a \textit{damping factor} of \begin{equation}
    e^{-\gamma} := (1 - 4p/3),
\end{equation}
while the identity is unchanged. 

For a tensor product channel acting on an $n$-qubit Pauli string
\begin{equation}
    P \equiv P_1 \otimes \cdots \otimes P_n \mapsto \mathcal{E}_p (P_1) \otimes \cdots \otimes \mathcal{E}_p (P_n) ,
\end{equation}
we see that
\begin{equation} \label{eqn:pauli-noise-mapping}
    P \equiv P_1 \otimes \cdots \otimes P_n \mapsto e^{-\gamma w[P]} P
\end{equation}
where $w[P]$ is the weight (number of non-identity Paulis) of Pauli $P$. 


In Majorana propagation,
\begin{equation}
    \mathcal{O} = \sum_i c_i M_i
\end{equation}
where $M_i$ are $n$-qubit Majorana operators (often called Majorana monomials in this context) of the form
\begin{equation} \label{eqn:majorana}
    M = s \gamma_1^{b_1} \gamma_2^{b_2} \cdots \gamma_{2n - 1}^{b_{2n - 1}} \gamma_{2n}^{b_{2n}} .
\end{equation}
Here, $s \in \mathbb{C}$ is a phase, the Majorana operators are defined by
\begin{align} \label{eqn:majorana-operators-odd}
    \gamma_{2j - 1} &:= \left( \prod_{k = 1}^{j - 1} Z_k \right) X_j \\
    \gamma_{2j} &:= \left( \prod_{k = 1}^{j - 1} Z_k \right) Y_j , \label{eqn:majorana-operators-even}
\end{align}
and $b_i \in \{0, 1\}$. 

Under a tensor product of single-qubit channels
\begin{equation}
    \mathcal{E}(M) := \left( \mathcal{E}_1 \otimes \cdots \otimes \mathcal{E}_n \right) (M) ,
\end{equation}
we can express the Majorana $M$ in the Pauli basis $\mathcal{P}(M)$ via~\eqref{eqn:majorana-operators-odd} -~\eqref{eqn:majorana-operators-even} to obtain a similar mapping to~\eqref{eqn:pauli-noise-mapping}, namely
\begin{equation} \label{eqn:majorana-noise-mapping}
    \mathcal{E}_p \left( \mathcal{P}(M)_1 \right) \otimes \cdots \otimes \mathcal{E}_p \left( \mathcal{P}(M)_n \right) = e^{-\gamma w[ \mathcal{P}(M)]} M .
\end{equation}
The only difference relative to~\eqref{eqn:majorana-noise-mapping} is that in~\eqref{eqn:pauli-noise-mapping} we must use the \textit{Pauli weight} $w[\mathcal{P}(M)]$ instead of the Majorana weight $w[M]$. Since Majoranas are nonlocal, a single Majorana operator can have high Pauli weight, e.g. $w[\mathcal{P}(\gamma_{2n}) ] = n$, leading to different damping factors in noisy simulation.

\subsection{Surrogate/Symbolic Propagation}\label{sec:surrogate-propagation}
Consider a parameterized quantum circuit $\mathcal{C}(\Theta)$, an observable $\mathcal{O}$, 
and an initial state $\ket{\Psi_0}$. It is often of interest to compute the expectation value of 
the observable with respect to the evolved state as a function of the parameters $\Theta$: 
\begin{equation}
    \mathbb{E}[\mathcal{O}](\Theta) = \bra{\Psi_0}{\mathcal{C}(\Theta)^\dagger \mathcal{O} \mathcal{C}(\theta)}\ket{\Psi_0}
\end{equation}
This is particularly relevant for variational quantum algorithms and quantum machine learning applications. 
Although one can compute the expectation value for each parameter setting $\Theta$ using numerical propagation, this 
can become computationally expensive for large circuits and many parameter settings. The parameters 
do not affect the circuit structure, rather they only affect the coefficients of the back-propagated operator. 
Therefore, rather than numerically propagating the operator for each $\Theta$, one can \emph{symbolically} 
propagate the operator once to obtain
\begin{equation}
    f : \Theta \mapsto \mathbb{E}[\mathcal{O}](\Theta) \coloneq \sum_{i} c_i(\Theta) \Tr(P_i \rho_0)
\end{equation}
i.e. a function that maps the parameters to the expectation value of the observable. Terms that 
are structurally negligible can be pruned from the back-propagated operator, resulting in a
\emph{surrogate model} that can be evaluated for any parameter setting $\Theta$. \Propaq{} implements surrogate propagation
and provides an API for saving, loading, and evaluating surrogate models.
The details of implementation and usage are described in Sec.~\ref{sec:surrogate-propagation-usage}.
It is worth noting that surrogate propagation is significantly more expensive than numerical propagation, 
as a result of having to track the symbolic coefficient tree and perform symbolic algebra, and the method should be used when many energy evaluations are required for parameter tuning such that fast subsequent energy evaluation outweighs the initial cost, as shown in the example of Fig.~\ref{fig:surrogate-optimization}.

\section{API and Usage} \label{sec:api}

In this section, we provide a ``getting started'' guide with \Propaq{}, starting with installation and proceeding through most common uses. This section is designed to be self-contained and provide more details on the features highlighted in Sec.~\ref{sec:features}, as well as show additional features.

\subsection{Installation and Dependencies}\label{sec:installation}

\Propaq{} can be installed on Linux, Windows and Mac via PyPI using the following command: 

\begin{code}[shell, codecap={Installation of \Propaq{} via PyPI.}, label={lst:install}]
pip install propaq
\end{code}

\noindent \Propaq{} is natively compatible with Qiskit circuits, and this is a dependency of the library. Compatibility with Cirq~\cite{larose_overview_2019}, FFSIM~\cite{sung_ffsim_2026}, OpenFermion~\cite{mcclean_openfermion:_2017}, and Quimb~\cite{gray_quimb_2018} (for hybrid Schr\"{o}dinger-Heisenberg simulation with matrix product states) are optional extras that can be installed via the following command:
\begin{code}[shell, codecap={Installation of \Propaq{} with optional extras.}, label={lst:install-extras}]
pip install propaq[cirq,ffsim,openfermion,hybrid]
\end{code}
\noindent The source code for \Propaq{} is available on GitHub~\cite{belagali_hkbelagalipropaq_2026}.

\subsubsection{Building for a specific CPU}

The wheels distributed on PyPI are built to be generic and compatible with a wide range of CPUs across different architectures.
Several hot paths in the propagation engine can be autovectorized by the compiler, and therefore benefit from CPU-specific optimizations. 
Therefore, for users who want to maximize performance on known CPU architecture, \Propaq{} can be built from source 
using a Rust toolchain:
\begin{code}[shell, codecap={Building \Propaq{} for the host CPU.}, label={lst:build-native}]
git clone https://github.com/hkbelagali/propaq
cd propaq
maturin build --release
pip install target/wheels/propaq-*.whl
\end{code}
\noindent This build will generally only work on the exact CPU architecture it was built on, therefore building in cluster environments must be done on the same target node that will be used for running the simulations.

\subsection{Basic Usage}

The following code block shows the basic usage of \Propaq{} to simulate a circuit. 
\begin{code}[language=python, codecap={Basic usage of \Propaq{}.}, label={lst:basic-usage}]
import qiskit

from propaq import PauliCircuit, PauliTermSum, PauliPropagator, UniformNoiseModel, CoefficientTruncator

qc: qiskit.QuantumCircuit = ...
obs: qiskit.SparsePauliOp = ...
initial_state: int = ...

# Convert to Propaq circuit and observable.
pc = PauliCircuit.from_qiskit(qc)
pts = PauliTermSum.from_sparse_pauli_op(obs)

# Construct a propagator.
prop = PauliPropagator(
    noise=UniformNoiseModel(damping=0.001), 
    truncation=[CoefficientTruncator(1e-6)],
    n_threads=16,
    progress_bar=True
    )

# Compute the expectation value.
result = prop.expectation_value(
    observable=pts, 
    circuit=pc, 
    initial_state=initial_state
)
result.expectation_value  # Type: float
\end{code}
\noindent  Here, we assume that the user has 
a quantum circuit defined as a \inlinecode{qiskit.QuantumCircuit}, an observable 
expressed as a \inlinecode{qiskit.SparsePauliOp}, and an initial bitstring represented as an integer (e.g. 0 for the all $|0\rangle^{\otimes n}$ state in the computational basis). After, a \inlinecode{PauliPropagator} is constructed with various options for noise, truncation, number of threads to use, and whether or not to show simulation progress. These options are used when the \inlinecode{expectation_value} method is called to compute the expectation value of the observable in the circuit and initial state. 

One can use the analogous method \inlinecode{from_cirq} to parse a
Cirq circuit. Additionally, the \inlinecode{MajoranaTermSum}, \inlinecode{MajoranaCircuit},
and \inlinecode{MajoranaPropagator}, along with their conversion methods from Qiskit, Cirq, OpenFermion, and ffsim can be used to perform Majorana propagation.
Note that these require optional dependencies as described in Sec.~\ref{sec:installation}.

\subsection{Truncation and Noise Policies}
\label{sec:truncation}

\subsubsection{Built-in Truncation Policies}
\Propaq{} implements truncation policies which can be used to control the growth of the operator during propagation. The built-in
truncation policies include \inlinecode{CoefficientTruncator}, \inlinecode{WeightTruncator} (numeric), and \inlinecode{FrequencyTruncator} (symbolic). The propagator objects accept a list of truncation policies, which are applied in the order provided. Users can also specify a \inlinecode{TermBudget}, which suppresses all other lossy truncation policies until the operator has at least the given number of live terms, keeping propagation exact until it has had room to grow. For example, the following list of truncation policies says to truncate terms with coefficients below $10^{-6}$ and weight above 5, but keep propagation exact below 100 terms.
\begin{code}[language=python, codecap={Truncation policies in \Propaq{}.}, label={lst:truncation-policies}]
truncation = [
    CoefficientTruncator(1e-6),
    WeightTruncator(5),
    TermBudget(min_terms=100),
]
\end{code}
\noindent This list of truncation policies can then be provided to a propagator, e.g. the \inlinecode{PauliPropagator} from Listing~\ref{lst:basic-usage}, and then used in simulation.

\subsubsection{Built-in Noise Models} 
\Propaq{} provides uniform depolarizing noise for both Pauli and Majorana propagation through the \inlinecode{UniformNoiseModel} class. 
This applies a uniform depolarizing channel to each qubit after every layer in the circuit.
\begin{code}[language=python, codecap={Uniform depolarizing noise.}, label={lst:uniform-noise}]
from propaq import UniformNoiseModel

# Generate a model scaling each term by exp(-0.001 * weight[P])
noise_model = UniformNoiseModel(damping=0.001)

# Pass into a Pauli/Majorana propagator
prop = PauliPropagator(noise=noise_model)
\end{code}
Since the propagation engine is generic, the \inlinecode{UniformNoiseModel} can be used with any propagator. For example, it can be used 
directly with Weyl-Heisenberg propagation (see Appendix~\ref{sec:weyl-propagation}) and naturally generalizes to a qudit depolarizing channel.

\subsubsection{Custom Noise and Truncation Policies} 

While \Propaq{} contains native support for depolarizing noise, custom noise models are supported. This can be done by inheriting from \inlinecode{GateNoiseModel}, as shown below. In this example, we implement a channel for depolarizing noise
\begin{equation}
    \mathcal{E}(\rho) = (1 - p) \rho + p Z \rho Z
\end{equation}
with $p = (1 - e^{-\gamma}) / 2$.

\begin{code}[language=python, codecap={Custom noise in \Propaq{}.}, label={lst:dephasing-noise}]
import math
from propaq.noise import GateNoiseModel

class DephasingGateNoise(GateNoiseModel):
    X_MASK = 0x5555555555555555

    def __init__(self, gamma: float) -> None: 
        self.gamma = gamma 
    
    def damping_factor_term(self,
        basis_kind: int,
        words: list[int], 
        n_units: int, 
        weight: int
    ) -> float:
        x_count = sum(
            bin(w & self.X_MASK).count("1") 
            for w in words
        ) 
        return math.exp(-self.gamma * x_count)
\end{code}

\noindent As shown, the necessary method to implement is \inlinecode{damping_factor_term} which returns the numerical value to damp coefficients of backpropagated observables by, depending on the basis and observable. Other custom channels can be defined in this manner and subsequently used with propagators in the normal way.

Since the backend of \Propaq{} is in Rust, custom noise in Python can be sped up by defining them directly in Rust. For this reason, \Propaq{} features a plugin application binary interface (ABI), 
through which users can implement their own noise and truncation policies in Rust, C, or ahead-of-time (AOT) compiled Julia.
This method works by accepting a shared library (\inlinecode{.so} on Linux, \inlinecode{.dylib} on Mac, and \inlinecode{.dll} on Windows) that specifies its permissions and implements the noise or truncation policy.
As an example, we demonstrate how to implement a noise plugin in C in Appendix~\ref{sec:custom-noise-in-c}, which defines \inlinecode{uniform_noise.c}. 
After this, the shared library is built so that it is available as \inlinecode{libuniform_noise.so} Then, it can be loaded and used in \Propaq{} as follows: 

\begin{code}[language=python, codecap={Using a custom noise plugin in \Propaq{}.}, label={lst:custom-noise-usage}]
from propaq.noise import NativeNoiseModel
from propaq.truncation import NativeTruncator

damping = 0.001 

# Load the shared library, and pass in damping as a JSON string
noise_model = NativeNoiseModel(
    'libuniform_noise.so', 
    config=f'{{"damping": {damping}}}'
)

# Use like any other noise model 
prop = PauliPropagator(noise=noise_model)

# Custom truncation policy
truncator = NativeTruncator(
    'libtruncator.so',
    config=f'{{"...": ...}}'
)

prop = MajoranaPropagator(truncation=truncator)
\end{code}

We remark that \Propaq{} employs a permission system for its noise and truncation plugins. This ensures that the plugin has access to 
the necessary data and resources, while allowing the engine to perform optimizations and parallelization depending on the access level.
For example, the action of a noise plugin that only requires the weight (such as uniform depolarizing noise) can be precomputed 
once at the beginning of the propagation. This virtually eliminates all overhead of the plugin, and allows for efficient 
parallel execution. In contrast, a plugin that requires access to full term's structure generally cannot be precomputed, and 
requires careful design to prevent race conditions and ensure reproducibility. The permissions and details for implementing plugins 
are described in the \Propaq{} documentation~\cite{belagali_hkbelagalipropaq_2026}.


\subsection{Arbitrary basis propagation} \label{sec:arbitrary-basis-backpropagation}
\Propaq{} supports propagation in arbitrary bases through its Python API. The user can define a custom basis 
representation and algebraic relations by subclassing the \inlinecode{AbstractTerm}, \inlinecode{DictTermSum}, and \inlinecode{AbstractPropagator} classes. This 
allows the user to utilize extrapolation methods and term I/O for eligible custom bases. As an example, we
study a single spin-$S$ system, which lives in a Hilbert space of dimension $d=2S+1$ with commutation relation
\begin{equation}
    [S_a, S_b] = i \epsilon_{abc} S_c  .
\end{equation}
For $S=1/2$, this reduces to the Pauli algebra. Note that this is not a complete 
operator basis for $S>1/2$. We can define a single parameterized gate $U_{\mathbf{n}}(\theta)$ 
for a rotation about an arbitrary axis $\mathbf{n}$, given by 
\begin{equation}
U_{\mathbf{n}}(\theta) = \exp{-i\theta (\mathbf{n} \cdot \vec S)}
\end{equation}
Then, via Wigner's theorem, we get the closed-form for every rotation $S$, 
\begin{equation} 
    U_{\mathbf{n}}^\dagger (\theta) S_a U_{\mathbf{n}}(\theta) = \sum_b R_{ab}(\mathbf{n}, \theta) S_b
\end{equation}
where $R_{ab}(\mathbf{n}, \theta) \in \mathrm{SO}(3)$ is a rotation matrix given by 
\begin{equation}
    R(\mathbf{n}, \theta) = \cos(\theta)I + \sin(\theta)[\mathbf{n}]_{\times}  + (1 - \cos(\theta))\mathbf{n}\mathbf{n}^\top
\end{equation}
where $[\mathbf{n}]_{\times}$ is the cross-product matrix of $\mathbf{n}$.
The overlap between a spin coherent state $\ket{\mathbf{n}} = U_{\mathbf{n}}(\theta)\ket{S, S}$ and
a spin operator $S_a$ can be computed as
\begin{equation}
    \bra{\mathbf{n}} S_a \ket{\mathbf{n}} = S \cdot n_a
\end{equation}

In the following code snippet, 
we demonstrate how to implement these spin operators in \Propaq{} by subclassing the appropriate methods.
\begin{code}[language=python, codecap={Subclassing \Propaq{} for a single spin-$S$ particle.}, label={lst:spin-basis}]
from dataclasses import dataclass
import numpy as np
from propaq.datatypes import AbstractTerm, DictTermSum
from propaq.propagators import AbstractPropagator

@dataclass(frozen=True, slots=True)
class SpinTerm(AbstractTerm):
    axis: int  # 0, 1, 2 for $S_x, S_y, S_z$
    spin: float  # $S$

    @property
    def weight(self) -> int: # 3. from (@\ref{subsec:arbitrary-basis}@)
        return 1

    @property
    def n_units(self) -> int:
        return 1

    def commutes_with(self, other: "SpinTerm") -> bool: # 2. from (@\ref{subsec:arbitrary-basis}@)
        return self.axis == other.axis

    def __matmul__(self, other: "SpinTerm") -> tuple[complex, "SpinTerm"]: # 1. from (@\ref{subsec:arbitrary-basis}@)
        raise NotImplementedError

    def trace_with_diag_state(self, diag_state) -> complex: # 4. from (@\ref{subsec:arbitrary-basis}@)
        return self.spin * diag_state[self.axis]

    def to_bytes(self) -> bytes: # 6. from (@\ref{subsec:arbitrary-basis}@)
        return bytes([self.axis, round(2 * self.spin)])

    @classmethod
    def from_bytes(cls, data: bytes, n_units: int) -> "SpinTerm": # 6. from (@\ref{subsec:arbitrary-basis}@)
        axis, doubled_spin = data[0], data[1]
        return cls(axis, doubled_spin / 2)

    def __hash__(self) -> int: # 6. from (@\ref{subsec:arbitrary-basis}@)
        return hash((self.axis, self.spin))

    def __eq__(self, other: object) -> bool: # 6. from (@\ref{subsec:arbitrary-basis}@)
        return isinstance(other, SpinTerm) and self.axis == other.axis and self.spin == other.spin

class SpinTermSum(DictTermSum[SpinTerm]):
    term_type = SpinTerm  # enables file I/O

class SpinRotation:
    def __init__(self, axis: tuple[float, float, float], angle: float):
        n = np.asarray(axis) / np.linalg.norm(axis)
        cross = np.array([[0, -n[2], n[1]], [n[2], 0, -n[0]], [-n[1], n[0], 0]])
        self.R = np.cos(angle) * np.eye(3) + np.sin(angle) * cross + (1 - np.cos(angle)) * np.outer(n, n)

class SpinPropagator(AbstractPropagator):
    term_sum_type = SpinTermSum

    def apply_gate(self, term: SpinTerm, coeff: complex, rotation: "SpinRotation"): # 5. from (@\ref{subsec:arbitrary-basis}@)
        for b in range(3):
            weight = rotation.R[term.axis, b]
            if weight:
                yield SpinTerm(b, term.spin), coeff * weight
\end{code}
Although the matrix multiplication method is not required for the spin basis, it is necessary to implement 
to satisfy the \inlinecode{AbstractTerm} interface. We also note that custom bases 
are not currently routed through the Rust backend, and are therefore slower than the built-in Pauli and Majorana bases by a non-trivial constant factor. We are actively working on extending support for custom bases into the Rust backend.

\subsection{Custom Gate Registration}
\begin{figure}
    \centering
    \begingroup
    \providecommand{\mathdefault}[1]{#1}
    \catcode`\_=12
    \resizebox{\columnwidth}{!}{\input{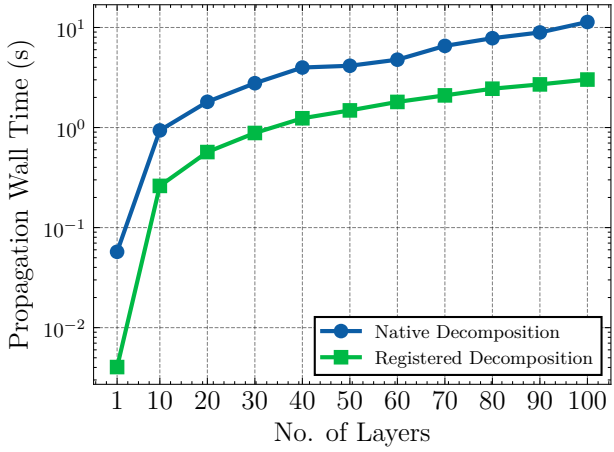}}
    \endgroup
    \caption{Propagation runtime for a 10-qubit non-Clifford brickwork exchange ansatz for a \Propaq{}'s native decomposition, and a custom decomposition registered with \Propaq{}.}
    \label{fig:gate-registry}
\end{figure}

Heisenberg propagation is a general procedure that can be applied to any gate, provided that it can be parameterized 
in terms of basis generators and angles. It is impractical to implement every possible gate in the core engine, therefore \Propaq{} 
decomposes gates into its native basis using quantum compilation techniques. Although this ensures that any gate can be accurately 
represented, the decompositions are not necessarily optimal, and can lead to larger-than-necessary operator growth. For cases 
where the user has a custom gate that is not in the native basis with a known, optimal decomposition, \Propaq{} allows 
the user to register the gate in terms of its generator representation. \Propaq{}
will validate the user-supplied generator representation by comparing it to its native decomposition. If the two representations are
equivalent, the gate can be used in the propagation engine in its optimized form.
As an example, we demonstrate how
to register a CX gate in \Propaq{} using its generator representation:
\begin{code}[language=python, codecap={Registering a CX gate in \Propaq{}.}, label={lst:register-gate}]
from propaq.circuits import PauliCircuit, pauli_rotation_generator, register_qiskit_gate

def cnot_terms(
    instr: qiskit.circuit.Instruction, 
    q_indices: tuple[int, int], 
    width: int, 
    rep: GateRep
    ): 
    i, j = q_indices # control, target
    n_qubits = rep.qubits_in_width(width)

    # Get the string repr of a Pauli
    def label(axis_i, axis_j): 
        chars = ["I"] * n_qubits
        if axis_i: 
            chars[n_qubits - 1 - i] = axis_i
        if axis_j:
            chars[n_qubits - 1 - j] = axis_j
        return "".join(chars)

    # Build the generator terms
    terms = [] 
    for axis_i, axis_j, coeff in (
        ("Z", None, math.pi / 2),
        (None, "X", math.pi / 2),
        ("Z", "X", -math.pi / 2),
    ):
        gen, unit = pauli_rotation_generator(
            rep, 
            label(axis_i, axis_j)
        )
        terms.append((gen, coeff * unit))
    return [terms]

# Register the qiskit gate
register_qiskit_gate("cx", cnot_terms)
\end{code}
\noindent For certain gates, gate registry can yield significant performance improvements. An example is shown in Fig.~\ref{fig:gate-registry} where custom decomposition registered with \Propaq{} achieves a 2-5x speedup when benchmarking the wall clock time of propagation for a ten-qubit non-Clifford brickwork exchange ansatz.

\subsection{Logging, Storage, and I/O}\label{sec:storage-logging-usage}

\subsubsection{Logging}

\Propaq{} supports logging of propagation statistics and runtime diagnostics. These features are provided by \inlinecode{Logger} objects. 

\begin{code}[language=python, codecap={Logging in \Propaq{}.}, label={lst:logging-usage}]
from propaq import Logger

# Save to propagation.log, and log every 5 gates.
logger = Logger(
    filename='propagation.log', 
    log_every=5,
)
prop = PauliPropagator(logger=logger)
\end{code}
\noindent This writes information to the log file at regular intervals defined by \inlinecode{log_every} in a JSONL format. The data can be parsed and analyzed using the 
\inlinecode{LogParser} class, which returns trajectories of each logged quantity. 
\begin{code}[language=python, codecap={Parsing logs in \Propaq{}.}, label={lst:log-parser-usage}]
from propaq import LogParser 

log = LogParser('propagation.log')
log.terms_discarded # list[int]
\end{code}
\noindent Additional logged quantities include \inlinecode{discarded_coeff_l1}, \inlinecode{discarded_coeff_max}, \inlinecode{ms_per_gate}, \inlinecode{elapsed_ms}, \inlinecode{gate_events},
 \inlinecode{terms_before}, \inlinecode{terms_after}, 
 and \inlinecode{truncation_events}, among others defined in \Propaq{}'s documentation~\cite{belagali_hkbelagalipropaq_2026}.

\subsubsection{Storage and I/O}

Storing propagated operators is a common requirement for many applications and downstream analysis. \Propaq{} provides a simple 
interface for storing and loading propagated operators in a binary format.
\begin{code}[language=python, codecap={Storing and loading propagated operators in \Propaq{}.}, label={lst:storage-usage}]
from propaq import PauliTermSum 

 # Propagate and save a term sum to disk.
prop = PauliPropagator(...) 
prop.propagate(..., filename='obs.gz')

# Load a term sum from disk.
termsum = PauliTermSum.from_file('obs.gz')
\end{code}

\noindent Term sums can also be saved to disk through via \inlinecode{termsum.save('obs.gz')}. These operations are \emph{eager} in memory, and therefore impractical for very large operators. 
Therefore, we also provide a lazy interface for loading and merging operators. This enables 
distributed workflows for propagation. Users can partition an observable and run on different devices, 
store the results in disk, and then merge the results into a single operator later for analysis.
\begin{code}[language=python, codecap={Lazy loading and merging of propagated operators in \Propaq{}.}, label={lst:lazy-storage-usage}]
from propaq import PauliTermStreamer

streamer = PauliTermStreamer.from_file('obs.gz') 

# Or: merge multiple operators into one 
files = ['obs1.gz', 'obs2.gz', 'obs3.gz']
obs = PauliTermSum() 
for f in files:
    obs.merge_from_file(
        PauliTermStreamer.from_file(f)
    )
\end{code}
\noindent Merging using a streamer ensures that the operator is not first materialized and then merged, which redundantly increases memory usage.

\subsection{Extrapolation Methods}\label{sec:extrapolation-usage}

\begin{figure}
    \centering
    \begingroup
    \providecommand{\mathdefault}[1]{#1}
    \catcode`\_=12
    \resizebox{\columnwidth}{!}{\input{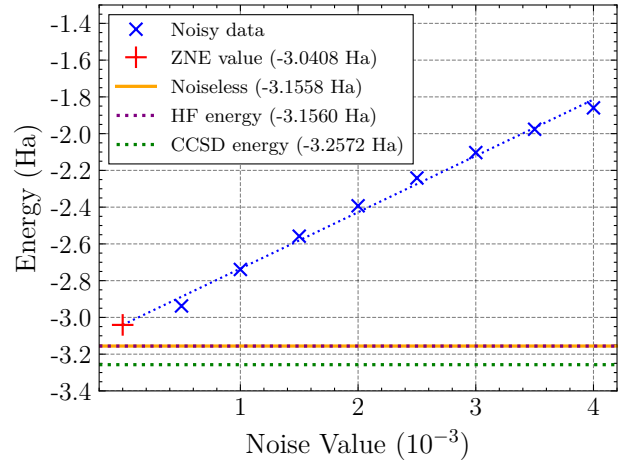}}
    \endgroup
    \caption{Zero-noise extrapolation of a 2-layer unitary cluster Jastrow (UCJ) ansatz for a $(\mathrm{H}_2)_3$ molecule. A uniform depolarizing 
    noise channel is applied after each layer of the ansatz, and a linear extrapolation is performed to estimate the expectation value.}
    \label{fig:zne}
\end{figure}

\Propaq{} provides classes and methods to implement extrapolation methods for zero-cutoff and zero-noise limit behavior of quantum circuits, which are detailed in the 
following section. \Propaq{} also allows users to implement extrapolators for custom noise and truncation policies by subclassing \inlinecode{ZeroNoiseExtrapolator} and 
\inlinecode{ZeroCutoffExtrapolator}, the details of which are mentioned in \Propaq{}'s documentation~\cite{belagali_hkbelagalipropaq_2026}.

\subsubsection{Zero-Cutoff Extrapolation}\label{sec:zce-usage}

An example result of zero-cutoff extrapolation improving the accuracy of expectation values was shown in Fig.~\ref{fig:zce}. The following code block shows the general structure for using zero-cutoff extrapolation in \Propaq{}.

\begin{code}[language=python, codecap={Zero-cutoff extrapolation in \Propaq{}.}, label={lst:zce-usage}]
from propaq import CoefficientCutoffExtrapolator
from propaq import CoefficientTruncator

circuit: PauliCircuit = ... 
observable: PauliTermSum = ...
prop: PauliPropagator = PauliPropagator(
    truncation=[CoefficientTruncator(1e-6)]
)

cce = CoefficientCutoffExtrapolator(
    fitting_fn = lambda x, a, b: a + b * x,
    cutoff_values=[1e-6, 1e-5, 1e-4, 1e-3]
)
result_cce = cce.run(
    propagator=prop,
    circuit=circuit,
    observable=observable,
    initial_state=0,
)
\end{code}
\noindent As shown, zero-cutoff extrapolation can be easily implemented by simply defining the fit function \inlinecode{fitting_fn} and set of cutoff values to run. We remark that \Propaq{} also defines a \inlinecode{WeightCutoffExtrapolator} which performs truncation based on observable weight. The usage for this extrapolator is identical to the above.

\subsubsection{Zero-Noise Extrapolation}\label{sec:zne-usage} 

In addition to zero-cutoff extrapolation, \Propaq{} also implements zero-noise extrapolation (ZNE). While ZNE has been implemented in several quantum error mitigation packages~\cite{larose_mitiq_2022,cirstoiu_volumetric_2023}, it is natural to consider ZNE for noisy Heisenberg propagation because, as discussed in Sec.~\ref{sec:background}, it reduces simulation complexity at larger noise levels. For this reason \Propaq{} natively implements ZNE through the \inlinecode{ZeroNoiseExtrapolator} class. An example of applying ZNE with \Propaq{} is shown below.

\begin{code}[language=python, codecap={Zero-noise extrapolation in \Propaq{}.}, label={lst:zne-usage}]
from propaq import ZeroNoiseExtrapolator

circuit: PauliCircuit = ... 
observable: PauliTermSum = ...
noise = UniformNoiseModel(damping=0.00)
prop = PauliPropagator(noise=noise)

zne = ZeroNoiseExtrapolator(
    fitting_fn = lambda x, a, b: a + b * x,
    noise_values=[0.005, 0.010, 0.015, 0.020]
)
result = zne.run(
    propagator=prop,
    circuit=circuit,
    observable=observable,
    initial_state=0,
)
\end{code}
\noindent Example results are shown in Fig.~\ref{fig:zne} for a two-layer unitary cluster Jastrow ansatz for a hydrogen chain. While both the bias and variance of ZNE is strongly noise and problem dependent, and still largely an active area of research, \Propaq{}'s implementation allows researchers to easily use and analyze these techniques in the context of Heisenberg propagation.

\subsection{Variational Algorithms}\label{sec:surrogate-propagation-usage}

Symbolic propagation allows for the construction of surrogate models for variational algorithms 
and parameter sweeps. The \inlinecode{SurrogatePropagator} class provides the interface 
necessary to run these algorithms, porting from Qiskit and Cirq objects.
\begin{code}[language=python, codecap={Surrogate propagation in \Propaq{}.}, label={lst:surrogate-usage}]
# Construct a parameterized Qiskit circuit
theta = ParameterVector('theta', 6) 
qc = QuantumCircuit(3) 

qc.x(0)
qc.x(2)

it = 0
for layer in range(2):
    for q in range(3): 
        qc.rz(theta[it], q) 
        it += 1

# Build the surrogate model
obs : PauliTermSum = ... 
circuit = SurrogatePauliCircuit.from_qiskit(qc)

model = PauliSurrogatePropagator().build(obs, circuit, initial_state=0)

# Wrap the model for ease of use
variational_model = VariationalSurrogateModel(
    model, circuit.parameter_sources, circuit.qiskit_parameters
)

# Optimize using scipy.minimize 
def cost(x): 
    return variational_model.evaluate(x)

result = minimize(cost, x0=np.random.rand(6), method='COBYLA', options={'maxiter': 1000})
\end{code}

\subsection{Hybrid Simulation}\label{sec:schrodinger-heisenberg-usage}
The hybrid Schr\"odinger--Heisenberg simulation is provided through the \inlinecode{hybrid_expectation_value} method. We provide a 
code snippet demonstrating how to use this method, assuming the user has a \inlinecode{quimb} \inlinecode{MatrixProductState} object, and a \inlinecode{PauliTermSum} 
back-propagated observable.
\begin{code}[language=python, codecap={Hybrid Schr\"odinger--Heisenberg simulation in \Propaq{}.}, label={lst:hybrid-usage}]
from quimb.tensor import MatrixProductState
from propaq import PauliTermSum, hybrid_expectation_value

# Forward-evolved state.
mps: MatrixProductState = ...

# Backward-evolved observable.
observable: PauliTermSum = ...

# Compute the expectation value 
ev = hybrid_expectation_value(observable, mps)
\end{code}
\noindent Here, the \inlinecode{mps} represents the forward simulation of the initial state to the circuit bipartation, and the \inlinecode{obs} represents the backward simulation of the observable to the bipartition. Of course, backpropagating \inlinecode{obs} to this point can be done with any propagation methods in \Propaq{} previously described.

\section{Design and Implementation} \label{sec:design}

\Propaq{} is implemented in Rust, with rayon used for multithreading and pyo3 used to expose the Rust API to Python. 
This section describes the design and implementation of \Propaq{}, including its data types and parallelization.
A general trend in the design of \Propaq{} is to use different data types for the Python frontend and the Rust backend,
with fast conversions between the two. This gives \Propaq{} a simple Python interface, and makes it compatible with most major quantum software packages written in Python, in addition to ensuring high performance.
Python and Rust differ in their design patterns, necessitating different data types and 
interfaces for the two languages, described in this section. We remark that several design decisions were adopted from \MonoProp{}~\cite{monoprop-github,monoprep} --- notably \inlinecode{BasisString}, \inlinecode{OperatorIndex}, and \inlinecode{InvertedIndex} data types, as well as partitioning, thread assignment, and pinning --- while other design elements, previously mentioned features, and the Rust backend/Python frontend, were uniquely chosen.

\subsection{Term Storage} 
\label{sec:term-storage}
The nature of Heisenberg propagation demands a data structure that can compactly represent enormous numbers of operators, and allow for efficient computation of their algebraic relations.
The symplectic representation of Pauli and Majorana operators is a natural choice, where the operators are represented as bitmasks, and the commutation relations become bitwise operations. 
Consistent with Rust's design philosophy, we separate the term storage (\inlinecode{BasisString}) from term behavior (\inlinecode{PauliAlgebra}/\inlinecode{MajoranaAlgebra}). 
This allows the engine to be generic over the basis, and promotes the implementation of new bases in the future.
\Propaq{} utilizes three main data types for representing operators: \inlinecode{Bitset}, \inlinecode{BasisString}, 
and \inlinecode{OperatorIndex}. The second two are 
integrated with the dynamically-sized \inlinecode{Bitset} data type for pyo3 boundary crossing.

\subsubsection{Bitset}
We implement a \inlinecode{Bitset} data type that can represent an arbitrary number of bits, and provide efficient implementations of the bitwise operations required for Heisenberg propagation.
The \inlinecode{Bitset} data type is a contiguous array of 64-bit unsigned integers, where each bit represents a mode (qubit or Majorana mode). 
It allocates 4 integers by default, which supports up to 256/128 qubits for Pauli/Majorana operators without heap allocation. 
\subsubsection{BasisString<W>}
The dynamic size of \inlinecode{Bitset} makes it convenient for traveling across the 
pyo3 boundary. However, since the width of the bitmask is unknown at compile time, the compiler is limited in the 
optimizations it can perform on the bitwise operations, which is important for hot loop operation performance. Therefore, we implement a \inlinecode{BasisString} data type that dispatches 
fixed-size operator representations for the Rust backend based on the system size. For example,
if the system has 14 qubits, the \inlinecode{BasisString<W>} will dispatch with $W = 1$ 64-bit word,
the next-largest available width. The qubit thresholds are 32, 64, 128, up to 2048 qubits/sites. Since the width is known at compile time, the compiler can perform
aggressive optimizations and autovectorization, resulting in performance improvements. \inlinecode{BasisString}, being a 
Rust const generic, cannot be used across the pyo3 boundary, and therefore its use in combination with 
\inlinecode{Bitset} allows for both performance and flexibility.

\subsubsection{OperatorIndex}
Although the \inlinecode{BasisString} data type is efficient for representing individual operators, it is 
not the most optimal way of representing many strings in a term sum. Note that each gate in a circuit 
will only affect a small number of qubits, and therefore for large systems, the majority of the bits 
in the bitmask will be zero. Additionally, propagation is commonly accompanied by weight truncation,
which will prune strings with a large number of non-zero bits. Therefore, in order to store 
many operators  $B_1, \cdots B_n$, we define the \inlinecode{OperatorIndex} 
data type which stores the indices of the non-trivial operations in a contiguous array for each key. 
This allows the propagator to easily identify which terms are affected by a gate. This architecture does not 
scale well if the number of non-trivial operations is large, and therefore \inlinecode{OperatorIndex} 
specifies an inline width $W$ after which it will fall back to storing \inlinecode{BasisString<W>} in a 
hash map. In the presence of weight truncation, the inline width is fixed to twice the weight cutoff 
so that overflow into the hashmap never occurs. Additionally, the engine 
collects statistics on the number of terms that overflow into the hashmap, and automatically 
increases the inline width when necessary, with a hard cap of 32.   

In addition to the \inlinecode{OperatorIndex} data type, we adopt the \inlinecode{InvertedIndex} 
view from \MonoProp{}~\cite{monoprop-github,monoprep}, which provides a transposed view of the rows. 
That is, rather than delivering the indices of non-trivial operations for each term, 
it delivers the terms that contain a non-trivial operation for each index. This allows for 
the efficient computation of anticommutation relations. For example, if the target gate's
generator acts on index $i, j, \ldots$, then the \inlinecode{InvertedIndex} can be used to compute the 
combined parity of the terms across these indices. This flags the terms that anticommute with the generator, 
whose bitmasks are materialized and passed downstream for computing the new terms. Thus, we are able 
to avoid materializing the bitmasks of all terms, leading to performance improvements for large term sums.

\subsection{Basis Algebra}
\label{sec:basis-algebra}
The types described in the previous section store the structure of the operators, but do not 
provide any interpretation of the algebraic relations between them. The \inlinecode{Basis} trait 
provides a contract for the algebraic relations of the basis, i.e. it informs the engine how to 
interpret a \inlinecode{BasisString<W>} in the context of the basis. The \inlinecode{Basis} trait is 
implemented for both the Pauli and Majorana bases in \inlinecode{PauliAlgebra} and \inlinecode{MajoranaAlgebra},
respectively.

\subsubsection{Pauli Strings}
A Pauli string is a tensor product of Pauli operators for each qubit~\eqref{eqn:pauli}.
It is often more convenient to represent a Pauli string in its symplectic representation, i.e. as a pair of bitmasks. 
For $n$ qubits, we define length $n$ bitmasks $x$ and $z$ such that 
\begin{equation}
    P = i^k \bigotimes_{i=1}^n X^{x_i} Z^{z_i}
\end{equation}
Therefore, in combination with a 64-bit floating point number to represent the coefficient, a Pauli string can be stored as a struct of \inlinecode{(coeff, x, z)} in exactly $64 + 2n$ bits. 
Consistent with the design of term storage, Pauli strings have a frontend (\inlinecode{PauliString}) and backend 
(\inlinecode{BasisString<W> + PauliAlgebra}).
For two Pauli strings $P_a = (x_a, z_a)$ and $P_b = (x_b, z_b)$ represented by their $x$ and $z$ bitmasks, we implement the product rule, commutation relations, operator weight, and overlap calculation with reference bitstrings, given as follows: 
\begin{equation}
P_a P_b = i^{\,k}\, P_c \qquad P_c \coloneq (x_a \oplus x_b, z_a \oplus z_b) 
\end{equation}
\begin{equation}
\{P_a, P_b\} = 0 \quad \text{iff} \quad p(x_a \cdot z_b) + p(z_a \cdot x_b) \mod 2 = 1 
\end{equation}
\begin{equation}
\text{weight}(P_a) = p(x_a \vee z_a) 
\end{equation}
\begin{equation}
\bra{\phi} P \ket{\phi} = \begin{cases} 
    0 & x \neq 0 \\ 
    (-1)^{p(z \cdot \phi)} & x = 0
\end{cases}
\end{equation}
where $p(\cdot)$ is the popcount function, $\ket{\phi}$ is a bitstring, $\oplus$ is the bitwise XOR operation, $\cdot$ is the bitwise AND operation, and $\vee$ is the bitwise OR operation. In addition to these algebraic relations, the data types also implement the necessary methods for serialization/deserialization, hashing, and equality comparison, enabling term I/O.
\subsubsection{Majorana Monomials} 
The $n$-qubit Majorana operators (often called \textit{Majorana monomials}), defined in~\eqref{eqn:majorana} and repeated here for convenience, are given by
\begin{equation}
    M = s\gamma_1^{b_1}\gamma_2^{b_2} \ldots \gamma_{2n-1}^{b_{2n-1}}\gamma_{2n}^{b_{2n}}
\end{equation}
where $s \in \mathbb{C}$ is a phase, $b_i$ are binary exponents, and $\gamma_i$ are the Majorana operators obeying the anticommutation relations.
\begin{equation}
    \gamma_i^\dagger = \gamma_i, \quad \{\gamma_i, \gamma_j\} = 2\delta_{ij}I 
\end{equation}
Therefore, the Majorana operators can similarly be represented as a $2n$-length bitmask, and the algebraic relations can be implemented as bitwise operations.
Similar to Pauli strings, Majorana monomials have a frontend (\inlinecode{MajoranaMonomial}) and backend (\inlinecode{BasisString<W> + MajoranaAlgebra}), with the following algebraic relations implemented for two Majorana monomial bitmasks $M_a = (a)$ and $M_b = (b)$, with $\ell_a$ and $\ell_b$ active modes: 
\begin{equation}\label{eq:hermiticity-phase} 
    s = i^{h(\ell_a)} \quad h(\ell) \coloneq \begin{cases} 
        0 & \ell \equiv 0, 1 \pmod 4 \\ 
        1 & \ell \equiv 2, 3 \pmod 4
    \end{cases}
\end{equation}
\begin{equation}\label{eq:majorana-product}
M_a M_b = i^{\,k}\, M_c \quad M_c \coloneq (a \oplus b)
\end{equation}
Here, $k$ is an integer determined by reordering of anticommuting Majoranas
and the phases $s_a,s_b,s_c$. 
\begin{equation}\label{eq:majorana-anticommute}
\{M_a, M_b\} = 0 \quad \text{iff} \quad \ell_a \ell_b + \ell_{a \cdot b} \equiv 1 \pmod 2
\end{equation}
\begin{equation}\label{eq:majorana-weight}
\text{weight}(M_a) = \text{weight}(\mathrm{JW}(M_a))
\end{equation}
\begin{equation}\label{eq:majorana-overlap} 
\bra{\phi} M_a \ket{\phi} = (-1)^{\lfloor n_{\text{paired}}/2\rfloor}\prod_{k \,:\, \text{paired \& active}} (2\phi_k - 1)
\end{equation} 
where $\mathrm{JW}(\cdot)$ is the Jordan-Wigner transformation, $\ket{\phi}$ is a bitstring/Slater determinant, and $s$ is the phase required to ensure the Hermiticity of the Majorana monomial. Note that this assumes that $a_{2k} = a_{2k - 1}$ for all $k$, otherwise $\bra{\phi} M_a \ket{\phi} = 0$. The weight is taken to be the \emph{Pauli weight} of the Jordan-Wigner-transformed monomial, in order to ensure consistency with noise models. The serialization/deserialization, hashing, and equality comparison methods are also implemented for Majorana monomials.

\subsection{Coefficient Representation}
\label{sec:coeff-repr}

\begin{figure*}
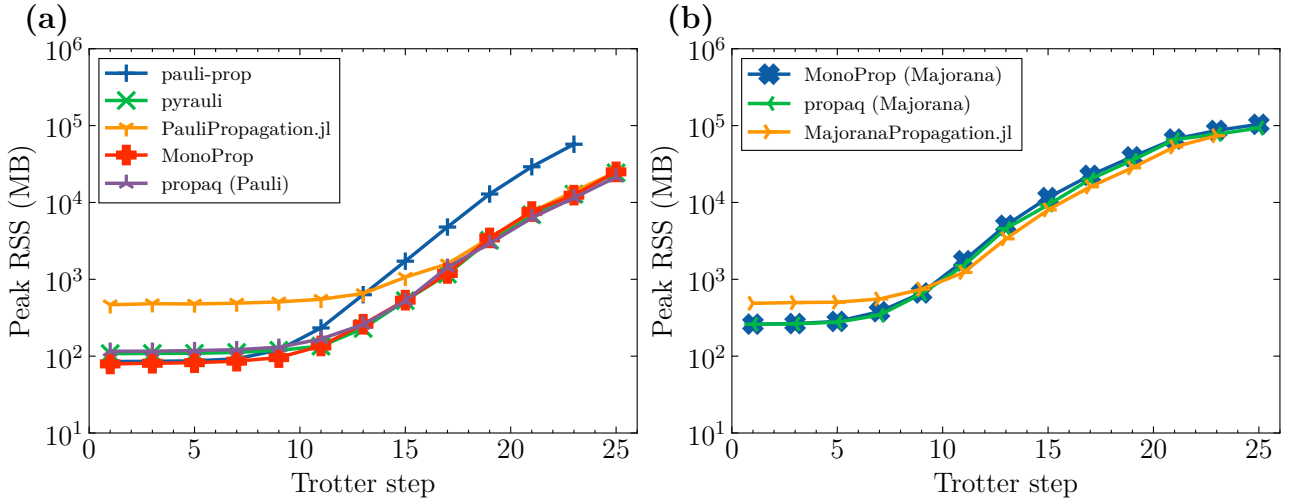

    \begingroup
    \providecommand{\mathdefault}[1]{#1}
    \catcode`\_=12
    \newcommand{\runtimepanel}[2]{%
        \begin{tikzpicture}
            \node[inner sep=0pt] (panelbox) {\resizebox{0.5\textwidth}{!}{\input{#2}}};
            \node[anchor=south west,font=\bfseries\large,inner sep=1pt] at ($(panelbox.north west)+(0.015\textwidth,-0.008\textwidth)$) {#1};
        \end{tikzpicture}%
    }
        \begin{tikzpicture}
            \node[inner sep=0pt] (panelbox) {\resizebox{0.5\textwidth}{!}{\input{figures/ising_memory.pgf}}};
            \node[anchor=south west,font=\bfseries\large,inner sep=1pt] at ($(panelbox.north west)+(0.015\textwidth,-0.008\textwidth)$) {(a)};
        \end{tikzpicture}%
        \begin{tikzpicture}
            \node[inner sep=0pt] (panelbox) {\resizebox{0.5\textwidth}{!}{\input{figures/hubbard_memory.pgf}}};
            \node[anchor=south west,font=\bfseries\large,inner sep=1pt] at ($(panelbox.north west)+(0.015\textwidth,-0.008\textwidth)$) {(b)};
        \end{tikzpicture}%
    
    \caption{Peak memory usage across various backends for computing expectation values of 
    (a) Single $Z$ observable trotterized time evolution of a 2D TFIM on a $6\times 6$ square lattice ($J = 1.0$, $h = 0.5$, $\Delta t = 0.1$)
    (b) Single $Z$ observable trotterized time evolution of a 2D Hubbard model on a $3 \times 3$ square lattice ($t = 1.0$, $U = 4.0$, $\Delta t = 0.1$). All backends use a coefficient cutoff of $10^{-6}$ and 64 threads, except for the single-threaded pauli-prop package.}
    \label{fig:ising-trotter-memory}
    \endgroup
\end{figure*}

The propagation engines are designed to be generic over the coefficient representation, allowing the 
same engine to be used for numeric propagation with different floating point precisions, as well as 
symbolic propagation. This functionality is provided by the \inlinecode{CoeffRepr} trait, which mainly 
wraps the coefficient type and provides a contract for the necessary algebraic operations. The numeric 
propagators currently support double (\inlinecode{f64}) and single (\inlinecode{f32}) precision floating point numbers.

The symbolic coefficients (\inlinecode{SymbolicCoeff}) are represented as directed acyclic graphs (DAGs) of algebraic operations, 
as is common in computer algebra systems (CAS). The arithmetic operations for scaling by a numerical coefficient,
adding two coefficients, and multiplying by a symbolic $\cos(\theta)$ or $\sin(\theta)$ term are 
all represented as nodes in the DAG. The nodes are wrapped in atomic reference-counted pointers (\inlinecode{Arc}) to 
allow for sharing of subexpressions across threads. Therefore, the symbolic propagation becomes a DAG construction, 
and the compilation to the final surrogate model walks the DAG to generate the final expression. The \inlinecode{Simplify} object identifies and merges 
shared symbolic histories. As one might expect, 
this is significantly more restricted than merging in a numerical propagator, and consequently symbolic 
propagation is more memory intensive and slower than numerical propagation.
\subsection{Propagation Engine}
\label{sec:engine}
The propagation engine is responsible for applying gates, noise channels, and truncation policies to the operator
in order to carry out the propagation. The observable is first divided over 
$n$ partitions, one for each thread. This enables multithreading via \inlinecode{rayon}. Additionally, the partitions, which disjointly divide the term space, are pinned to their corresponding spawner
threads on Linux systems to promote cache locality of the term sums. The circuit is applied in reverse order, and as a result the noise channels are applied prior to each layer.  \\ 
For each gate, the propagator first checks if it can be deferred through the 
Clifford deferral mechanism (Sec~\ref{sec:optimizations}). If so, its adjoint action is stored in a \inlinecode{CliffordTableau}. Once a non-Clifford gate is encountered later, it is conjugated through the tableau and applied to the term sum. Next, each owner thread identifies the terms affected by the gate through the \inlinecode{OperatorIndex}/\inlinecode{InvertedIndex} data structures described in Sec.~\ref{sec:term-storage}. 

From here, the affected terms' bitmasks in each partition are materialized by the owner thread. In the presence of coefficient truncation, the propagator estimates the coefficient magnitude of new terms and might choose not to produce them if they do not meet the threshold. For the new terms that are produced, they are then pruned according to given truncation policies. Terms not meeting the coefficient cutoff are stashed rather than being immediately dropped. It is common for an owner thread to produce terms not in its partition, and these terms must be routed to the correct owner thread. This is batched and done after each gate application across all threads to reduce the overhead of thread communication. 

In addition to this mechanism, if any term that was stashed due to truncation is later found in a partition, it is retrieved and added back to the term sum. Since the term is already being stored in memory, this rescue mechanism minimizes information loss while being free in memory. Note that terms dropped due to weight truncation are not rescued, since this truncation is structural. The scan over the terms during this phase is used to compute truncation statistics, which are logged to file if logging is enabled. After the routing, the noise channels are applied, and the next gate is processed. This continues until all gates have been applied, at which point either the final term sum is returned, or the expectation value is calculated by computing term overlaps with the initial state.

\section{Memory and Thread Performance} \label{sec:performance}

As shown in Fig.~\ref{fig:ising-trotter-runtime}, \Propaq{} meets or exceeds state-of-the-art Heisenberg propagation packages in terms of runtime. Here we show the analogous benchmark for memory. Specifically, we compare memory usage for the same benchmark in Fig.~\ref{fig:ising-trotter-runtime}, namely computing a Pauli $Z$ expectation value for two-dimensional $6 \times 6$ transverse-field Ising model and an $3 \times 3$ Hubbard model. These results are shown in Fig.~\ref{fig:ising-trotter-memory}. In these benchmarks, all simulators use the same coefficient cutoff of $10^{-6}$. As can be seen, despite the differences in runtime between packages in Fig.~\ref{fig:ising-trotter-runtime}, they are much more similar in terms of memory usage. In this case, we still see \Propaq{} performing on par or better than current state of the art, both in the Pauli and the Majorana basis. We expect to see the arbitrary basis support in \Propaq{} enable much more efficient simulations for certain applications, both in terms of memory and runtime, and seek to explore these applications in future work.

\begin{figure}
    \centering
    \begingroup
    \providecommand{\mathdefault}[1]{#1}
    \catcode`\_=12
    \resizebox{\columnwidth}{!}{\input{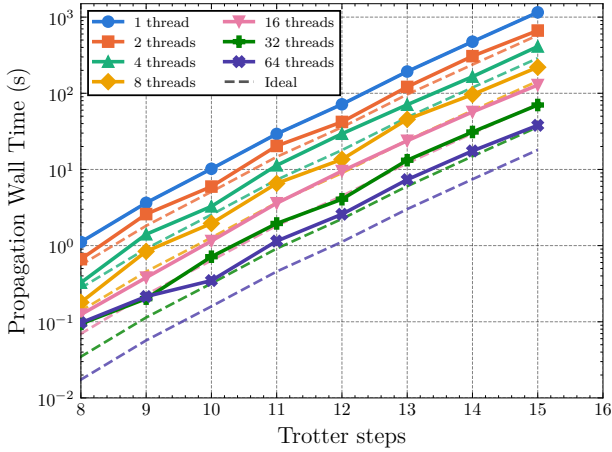}}
    \endgroup
    \caption{Propagation wall-clock time vs. Trotter step count for \Propaq{} running on 1--64 threads on a 2D $6 \times 6$ TFIM, compared to ideal linear scaling.}
    \label{fig:thread-scaling}
\end{figure}

Additionally, we study how performance scales with the number of threads. To do so, we record the wall-clock time to compute the energy of a two-dimensional $6 \times 6$ transverse-field Ising model vs the number of Trotter steps. These results, shown in Fig.~\ref{fig:thread-scaling}, demonstrate that multiple threads can result in significant speedups. Indeed, we roughly see a 10-20x speedup from one to 64 threads in this example. Additional threads on a single CPU, or on multiple CPUs, can further improve performance.

\section{Conclusion}

We introduced \Propaq{}, a highly general and flexible package for Heisenberg propagation. We showed that \Propaq{} performs as good or better than state-of-the-art Heisenberg simulation packages in terms of both runtime and memory. Because \Propaq{} contains multiple features that stand out from current software, we expect it to be useful for both quantum software developers and researchers. Notably, several of these features open the possibility for tackling open research questions, for example choosing the optimal basis for Heisenberg propagation and choosing the best circuit partitioning for hybrid Schr\"{o}dinger-Heisenberg simulation. We seek to tackle these open questions in future work enabled by \Propaq{}, and make it freely and openly available to the community for use.





\section*{Acknowledgments}

We acknowledge the use of generative AI for assistance writing some parts of the code in \Propaq{}, and for creating Fig.~\ref{fig:schrodinger-heisenberg-schematic}.
We acknowledge~\cite{SciencePlots} for making plots.

\bibliographystyle{unsrt}
\bibliography{references}

\onecolumngrid\newpage
\appendix

\section{Weyl-Heisenberg propagation}\label{sec:weyl-propagation}
In this section, we describe Heisenberg propagation in the Weyl-Heisenberg basis, 
which is a generalization of the Pauli basis to qudits. We provide the algebraic 
relations for the Weyl-Heisenberg operators and code snippets subclassing 
\Propaq{}'s classes to implement Weyl-Heisenberg propagation. \\\\ 
\subsection{Weyl-Heisenberg Algebra}
On one \emph{qudit} of dimension $d$, the Weyl-Heisenberg operators are defined by
the following relations:
\begin{equation}
    X\ket{j} \coloneq \ket{j + 1 \mod d}, \quad Z\ket{j} \coloneq \omega^j\ket{j}, \quad \omega \coloneq e^{2\pi i / d}
\end{equation}
obeying $X^d = Z^d = I$ and $ZX = \omega XZ$.
Then, a Weyl-Heisenberg string on $n$ qudits can be represented using two exponent vectors $a, b \in \mathbb{Z}_d^n$ as 
\begin{equation}
    W(a, b) \coloneq \tau^{a \cdot b} X^a Z^b, \quad X^a Z^b \coloneq \bigotimes_k X^{a_k} Z^{b_k}
\end{equation}
The prefactor $\tau$ is a phase factor, for which we will fix the convention 
\begin{equation}
    \tau \coloneq \begin{cases}
        \omega^{(d + 1) / 2} & d \text{ odd} \\
        e^{i\pi / d} & d \text{ even}
    \end{cases}
\end{equation}
Note that for $d = 2$, these relations reduce to the Pauli algebra.
Computing the product of two Weyl-Heisenberg strings $W(u)$ and 
$W(v)$ yields 
\begin{equation}\label{eq:weyl-product}
    W(v)\,W(u) \;=\; \tau^{\,E(v,u)}\; W\big(\bar{s},\bar{t}\big), \qquad
\begin{aligned} \bar{s} &= v_a + u_a \bmod d \\ \bar{t} &= v_b + u_b \bmod d \end{aligned}
\end{equation}
where the exponent $E(v, u)$ is given by
\begin{equation}\label{eq:weyl-exponent}
    E(v,u) \;=\; v_a\!\cdot\!v_b \;+\; u_a\!\cdot\!u_b \;+\; 2\,u_a\!\cdot\!v_b \;-\; \bar{s}\cdot\bar{t} \pmod{2d}    
\end{equation}
This means that the commutation relations of Weyl-Heisenberg strings are as follows: 
\begin{equation}\label{eq:weyl-symplectic}
    W(u)W(v) = \omega^{\sigma(v, u)} W(v)W(u) \qquad \sigma(v, u) = v_a\cdot u_b - u_a \cdot v_b \pmod{d} 
\end{equation}
where $\sigma$ is the symplectic form over $\mathbb{Z}_d$. 
Finally, we will need the overlap of a Weyl-Heisenberg string with a bitstring state 
$\ket{f} = \ket{f_1, f_2, \ldots, f_n}$, which is given by
\begin{equation}\label{eq:weyl-overlap}
    \bra{f} W(a, b) \ket{f} = \delta_{a, 0} \omega^{b \cdot f}
\end{equation}
We now have all the necessary relations to set up the term representation for Weyl-Heisenberg strings.
\begin{code}[language=python, codecap={Subclassing \Propaq{} to implement Weyl-Heisenberg strings.} , label={lst:weyl-heisenberg}]
from propaq.datatypes import AbstractTerm

def tau_pow(d: int, e: int) -> complex: 
    """
    Compute the phase factor $\tau^e$
    """ 
    # Determine k based on the parity of d
    k = 1 if d % 2 == 0 else d+1
    return cmath.exp(1j * math.pi * k * (e % (2*d)) / d) 

def omega_pow(d: int, e: int) -> complex:
    """
    Compute the phase factor $\omega^e$
    """ 
    return cmath.exp(2j * math.pi * (e % d) / d)

@dataclass(frozen=True)
class WeylString(AbstractTerm): 
    """$W(a, b) = \tau^{a \cdot b} X^a Z^b$"""

    a: tuple[int, ...] # exponent vector for $X$, $a \in \mathbb{Z}_d^n$
    b: tuple[int, ...] # exponent vector for $Z$, $b \in \mathbb{Z}_d^n$
    d: int

    @property
    def n_units(self) -> int: 
        return len(self.a) 
    
    @property 
    def weight(self) -> int: 
        return sum(1 for ai, bi in zip(self.a, self.b) if ai or bi)
    
    def __matmul__(self, other: "WeylString") -> tuple[complex, "WeylString"]: 
        d, e, na, nb = self.d, 0, [], []
        for va, vb, ua, ub in zip(self.a, self.b, other.a, other.b):
            s, t = (va + ua) % d, (vb + ub) % d
            e += va * vb + ua * ub + 2 * ua * vb - s * t # (@(Eq~\ref{eq:weyl-exponent})@)
            na.append(s)
            nb.append(t)
        return tau_pow(d, e), WeylString(tuple(na), tuple(nb), d)  # (@(Eq~\ref{eq:weyl-product})@)

    def dagger(self) -> tuple[complex, "WeylString"]: 
        inv = WeylString(
            tuple((-x) % self.d for x in self.a), tuple((-x) % self.d for x in self.b), self.d
        )
        phase, _ = self @ inv
        return 1 / phase, inv
    
    def symplectic(self, other: "WeylString") -> int: 
        return sum(
            va * ub - ua * vb for va, vb, ua, ub in zip(self.a, self.b, other.a, other.b)
        ) % self.d # (@(Eq~\ref{eq:weyl-symplectic})@)

    def commutes_with(self, other: "WeylString") -> bool:
        return self.symplectic(other) == 0

    def trace_with_diag_state(self, diag_state: int | tuple[int, ...]) -> complex:
        if any(self.a):
            return 0j
        if isinstance(diag_state, int):
            diag_state = tuple((diag_state // self.d**k) % self.d for k in range(self.n_units))
        return omega_pow(self.d, sum(bi * fi for bi, fi in zip(self.b, diag_state)))  # (@(Eq~\ref{eq:weyl-overlap})@)

    def to_bytes(self) -> bytes:
        return bytes(self.a) + bytes(self.b) + bytes([self.d])

    @classmethod
    def from_bytes(cls, data: bytes, n_units: int) -> "WeylString":
        a, b, d = data[:n_units], data[n_units : 2 * n_units], data[-1]
        return cls(tuple(a), tuple(b), d)

    def __hash__(self) -> int:
        return hash((self.a, self.b))

    def __eq__(self, other: object) -> bool:
        return (
            isinstance(other, WeylString) and self.a == other.a and self.b == other.b
        )

def identity(n: int, d: int) -> WeylString:
    return WeylString((0,) * n, (0,) * n, d)

def single(n: int, d: int, k: int, ai: int, bi: int) -> WeylString:
    """$X^{a_i} Z^{b_i}$ acting on site ``k`` alone."""
    a, b = [0] * n, [0] * n
    a[k], b[k] = ai % d, bi % d
    return WeylString(tuple(a), tuple(b), d)
\end{code}
\subsection{Weyl-Heisenberg Gates}
A Weyl-Heisenberg string is not generally Hermitian for $d > 2$. We'll consider only the 
Hermitian component of the Weyl-Heisenberg strings as unitary gate generators: 
\begin{equation}
    H_g = \frac12 (W_g + W_g^\dagger)\qquad U_g(\theta) = \exp(-i\frac{\theta}{2}H_g)
\end{equation}
Note that for $d = 2$, we have $W_g = W_g^\dagger$ and therefore $U = \exp(-i\theta W_g)$, which is the usual Pauli gate.
Because $W_g^d = I$, we can write the eigenprojectors of $W_g$ as
$\Pi_k = \frac{1}{d}\sum_m \omega^{-km} W_g^m$ for $k = 0, \ldots, d-1$. 
Then, we can write the unitary gate as a linear combination of the eigenprojectors, where the coefficients are given by:
 \begin{equation}\label{eq:weyl-coeffs}
U_g(\theta) = \sum_{m=0}^{d-1}c_m W_g^m, \qquad c_m = \frac{1}{d}\sum_{k=0}^{d-1} \exp{-i \frac{\theta}{2} \cos(2\pi k /d)} \omega^{-km}
\end{equation}
Once again, note that for $d=2$ we recover the familiar $c_0 = \cos(\theta/2)$ and $c_1 = -i\sin(\theta/2)$.
Let $s = \sigma(v, g)$, so  $W_g^j W(v)= \omega^{js} W(v) W_g^j \forall j$ as in Eq.~\eqref{eq:weyl-symplectic}.
We are now equipped to derive the propagation rule for a Weyl-Heisenberg gate's action: 
\begin{equation}
    U_g^\dagger W(v)U_g = \sum_{m,m'} \bar{c}_{m'}c_m W_g^{-m'} W(v) W_g^m = \sum_{m,m'}\bar{c}_{m'}c_m \omega^{-m's}W(v) W_g^{m-m'}
\end{equation}
and allowing $j = m - m'$, we have 
\begin{equation}\label{eq:weyl-branching-rule}
    U_g(\theta)^\dagger W(v) U_g(\theta) = \sum_{j=0}^{d-1} \Lambda_s[j] W(v) W_g^j \qquad \Lambda_s[j] = \sum_{m=0}^{d-1} \bar{c}_{m}c_{m+j}\omega^{-ms} 
\end{equation}
Therefore, the propagation of a Weyl-Heisenberg string through a gate can result in up to $d$ branches. 
$\Lambda$ is a $d \times d$ table dependent on the gate, and each term in the sum 
contributes its symplectic form $s$ to the phase factor. \\\\
Consider the equation for $d=2$. Then, we have $c_0 = \cos(\theta/2)$, and $c_1 = -i \sin(\theta/2)$.
Since we operate over $\mathbb{Z}_2$, we have $s=0$, or $s=1$. When $s=0$, we have 
\begin{align}
    \Lambda_0[0] &= \bar{c}_0 c_0 + \bar{c}_1 c_1 = \cos^2(\theta/2) + \sin^2(\theta/2) = 1 \\ 
    \Lambda_0[1] &= \bar{c}_0 c_1 + \bar{c}_1 c_0 = -i\cos(\theta/2)\sin(\theta/2) + i\sin(\theta/2)\cos(\theta/2) = 0
\intertext{When $s=1$, we have} 
    \Lambda_1[0] &= \cos^{2}(\theta/2) - \sin^2(\theta/2) = \cos(\theta) \\
    \Lambda_1[1] &= -i\cos(\theta/2)\sin(\theta/2) - i\sin(\theta/2)\cos(\theta/2) = -i\sin(\theta) 
\end{align}
This corresponds exactly to the Pauli propagation rule. We can now implement the Weyl-Heisenberg generator parameterization:
\clearpage
\begin{code}[language=python, codecap={Weyl-Heisenberg generator parameterization.}, label={lst:weyl-generator}]
@lru_cache(maxsize=None)
def _gate_coefficients(d: int, theta: float) -> tuple[complex, ...]:
    """$c_m$ such that $U_g(\theta) = \sum_m c_m W_g^m$."""
    k = np.arange(d)
    eigs = np.exp(-0.5j * theta * np.cos(2 * np.pi * k / d))
    return tuple(np.sum(eigs * np.exp(-2j * np.pi * k * m / d)) / d for m in range(d)) # (@(Eq~\ref{eq:weyl-coeffs})@)

class WeylRotation:
    """
    A unitary gate generated by the Hermitian component of a Weyl string.
    """

    def __init__(self, generator: WeylString, angle: float):
        d, n = generator.d, generator.n_units
        self.generator, self.angle = generator, angle

        powers = [(1 + 0j, identity(n, d))]
        for _ in range(d - 1):
            phase, key = powers[-1]
            step_phase, step_key = key @ generator
            powers.append((phase * step_phase, step_key))
        phase, key = powers[-1]
        step_phase, step_key = key @ generator
        if step_key != identity(n, d) or abs(phase * step_phase - 1) > 1e-12:
            raise ValueError("generator does not satisfy W^d = 1")
        self.powers = powers

        c = np.asarray(_gate_coefficients(d, angle))
        omega = np.array([[omega_pow(d, -m * s) for m in range(d)] for s in range(d)])

        # $\Lambda_s[j]$, (@(Eq~\ref{eq:weyl-branching-rule})@)
        self.table = np.array(
            [[np.sum(np.conj(c) * np.roll(c, -j) * omega[s]) for j in range(d)] for s in range(d)]
        ) 

def conjugate(key: WeylString, coeff: complex, rot: WeylRotation):
    """Yield the (child, coefficient) pairs of one term under one gate."""
    row = rot.table[key.symplectic(rot.generator)]  
    for j in range(key.d):
        weight = row[j]
        if weight == 0:
            continue
        power_phase, power_key = rot.powers[j]
        phase, child = key @ power_key
        yield child, coeff * weight * power_phase * phase
\end{code} 
Now, we have everything required to implement the Weyl-Heisenberg propagator. We will subclass 
\inlinecode{AbstractPropagator} and write small wrappers around the truncation and noise policies to 
make them compatible with the Weyl-Heisenberg term representation.
\begin{code}[language=python, codecap={Subclassing \Propaq{} to implement Weyl-Heisenberg propagation.}, label={lst:weyl-propagator}]
from propaq.propagators import AbstractPropagator
from propaq.datatypes import DictTermSum

class WeylTermSum(DictTermSum[WeylString]):
    term_type = WeylString # enables file I/O

class WeylPropagator(AbstractPropagator): 
    """Heisenberg propagation of Weyl-Heisenberg observables through qudit circuits."""

    term_sum_type = WeylTermSum

    def apply_gate(self, term: WeylString, coeff: complex, rotation: "WeylRotation"): 
        """Yield the (child, coefficient) pairs of one term under one gate.""" 
        return conjugate(term, coeff, rotation)
\end{code}
We have provided an example implementation of this Weyl-Heisenberg propagator on \Propaq's website, extensively validating the 
implementation against density matrix propagation and validating against known analytical results for the Potts model. We also 
demonstrate the out-of-box use of zero noise extrapolation. This can be 
found at \url{https://hkbelagali.github.io/propaq/examples/usage/11_custom_basis_qudit_weyl/}.

\section{Custom noise plugin implementation in C}  \label{sec:custom-noise-in-c}
\begin{code}[language=c, width=\textwidth, codecap={Implementing uniform depolarizing noise in C for \Propaq{}.}, label={lst:custom-noise}]
// uniform_noise.c
#include <math.h> 
#include <stdint.h> 
#include <stdlib.h> 
#include <string.h> 

#define PROPAQ_NOISE_ABI_VERSION 1u // ABI version 1.0

typedef struct { 
    double damping; 
} Ctx;

uint32_t propaq_noise_abi_version(void) { 
    return PROPAQ_NOISE_ABI_VERSION; // Return the ABI version of the plugin for compatibility checks 
}

// Parse a JSON string to extract the model parameters
static double parse_damping(const char* config_json) { 
    if (config_json == NULL) return 0.0; // Default damping if no config is provided
    const char* key = strstr(config_json, "\"damping\"");

    if (key == NULL) return 0.0; // Default damping if key not found

    const char* colon = strchr(key, ':');
    if (colon == NULL) return 0.0; 
    return atof(colon + 1); 
}

// Create a new noise model context
void* propaq_noise_create(const char* config_json) { 
    Ctx* ctx = (Ctx*)malloc(sizeof(Ctx)); 
    ctx->damping = parse_damping(config_json); 
    return ctx;
}

// Free alloc for model
void propaq_noise_destroy(void* ctx) { free(ctx); }

// Implement the noise application 
// Args:
//  ctx: pointer to the noise model context
//  basis_kind: the enum for the basis (Pauli, Majorana, etc.)
//  words: pointer to the array of strings
//  n_words: number of words in the array
//  n_units: number of units (qubits, fermions, etc.)
//  weight: the weight of the operator
//  layer_index: the index of the current layer in the circuit
//  n_layers: the total number of layers in the circuit
double propaq_noise_factor(void* ctx, uint32_t basis_kind, const uint64_t* words, size_t n_words,
                           uint32_t n_units, uint32_t weight, uint32_t layer_index,
                           uint32_t n_layers) {
    (void)basis_kind; (void)words; (void)n_words; (void)n_units;
    (void)layer_index; (void)n_layers;
    double damping = ((Ctx*)ctx)->damping;
    return exp(-damping * (double)weight); // exp(damping * w[P])
}
\end{code}
\end{document}

%% file: figures/surrogate.pgf
\begingroup%
\makeatletter%
\begin{pgfpicture}%
\pgfpathrectangle{\pgfpointorigin}{\pgfqpoint{4.127475in}{3.030312in}}%
\pgfusepath{use as bounding box, clip}%
\begin{pgfscope}%
\pgfsetbuttcap%
\pgfsetmiterjoin%
\definecolor{currentfill}{rgb}{1.000000,1.000000,1.000000}%
\pgfsetfillcolor{currentfill}%
\pgfsetlinewidth{0.000000pt}%
\definecolor{currentstroke}{rgb}{1.000000,1.000000,1.000000}%
\pgfsetstrokecolor{currentstroke}%
\pgfsetdash{}{0pt}%
\pgfpathmoveto{\pgfqpoint{0.000000in}{0.000000in}}%
\pgfpathlineto{\pgfqpoint{4.127475in}{0.000000in}}%
\pgfpathlineto{\pgfqpoint{4.127475in}{3.030312in}}%
\pgfpathlineto{\pgfqpoint{0.000000in}{3.030312in}}%
\pgfpathlineto{\pgfqpoint{0.000000in}{0.000000in}}%
\pgfpathclose%
\pgfusepath{fill}%
\end{pgfscope}%
\begin{pgfscope}%
\pgfsetbuttcap%
\pgfsetmiterjoin%
\definecolor{currentfill}{rgb}{1.000000,1.000000,1.000000}%
\pgfsetfillcolor{currentfill}%
\pgfsetlinewidth{0.000000pt}%
\definecolor{currentstroke}{rgb}{0.000000,0.000000,0.000000}%
\pgfsetstrokecolor{currentstroke}%
\pgfsetstrokeopacity{0.000000}%
\pgfsetdash{}{0pt}%
\pgfpathmoveto{\pgfqpoint{0.773727in}{0.279167in}}%
\pgfpathlineto{\pgfqpoint{4.077475in}{0.279167in}}%
\pgfpathlineto{\pgfqpoint{4.077475in}{2.980312in}}%
\pgfpathlineto{\pgfqpoint{0.773727in}{2.980312in}}%
\pgfpathlineto{\pgfqpoint{0.773727in}{0.279167in}}%
\pgfpathclose%
\pgfusepath{fill}%
\end{pgfscope}%
\begin{pgfscope}%
\pgfsetbuttcap%
\pgfsetroundjoin%
\definecolor{currentfill}{rgb}{0.000000,0.000000,0.000000}%
\pgfsetfillcolor{currentfill}%
\pgfsetlinewidth{0.501875pt}%
\definecolor{currentstroke}{rgb}{0.000000,0.000000,0.000000}%
\pgfsetstrokecolor{currentstroke}%
\pgfsetdash{}{0pt}%
\pgfsys@defobject{currentmarker}{\pgfqpoint{0.000000in}{0.000000in}}{\pgfqpoint{0.000000in}{0.041667in}}{%
\pgfpathmoveto{\pgfqpoint{0.000000in}{0.000000in}}%
\pgfpathlineto{\pgfqpoint{0.000000in}{0.041667in}}%
\pgfusepath{stroke,fill}%
}%
\begin{pgfscope}%
\pgfsys@transformshift{1.292239in}{0.279167in}%
\pgfsys@useobject{currentmarker}{}%
\end{pgfscope}%
\end{pgfscope}%
\begin{pgfscope}%
\pgfsetbuttcap%
\pgfsetroundjoin%
\definecolor{currentfill}{rgb}{0.000000,0.000000,0.000000}%
\pgfsetfillcolor{currentfill}%
\pgfsetlinewidth{0.501875pt}%
\definecolor{currentstroke}{rgb}{0.000000,0.000000,0.000000}%
\pgfsetstrokecolor{currentstroke}%
\pgfsetdash{}{0pt}%
\pgfsys@defobject{currentmarker}{\pgfqpoint{0.000000in}{-0.041667in}}{\pgfqpoint{0.000000in}{0.000000in}}{%
\pgfpathmoveto{\pgfqpoint{0.000000in}{0.000000in}}%
\pgfpathlineto{\pgfqpoint{0.000000in}{-0.041667in}}%
\pgfusepath{stroke,fill}%
}%
\begin{pgfscope}%
\pgfsys@transformshift{1.292239in}{2.980312in}%
\pgfsys@useobject{currentmarker}{}%
\end{pgfscope}%
\end{pgfscope}%
\begin{pgfscope}%
\definecolor{textcolor}{rgb}{0.000000,0.000000,0.000000}%
\pgfsetstrokecolor{textcolor}%
\pgfsetfillcolor{textcolor}%
\pgftext[x=1.292239in,y=0.230556in,,top]{\color{textcolor}{\rmfamily\fontsize{13.000000}{15.600000}\selectfont\catcode`\^=\active\def^{\ifmmode\sp\else\^{}\fi}\catcode`\%=\active\def
\end{pgfscope}%
\begin{pgfscope}%
\pgfsetbuttcap%
\pgfsetroundjoin%
\definecolor{currentfill}{rgb}{0.000000,0.000000,0.000000}%
\pgfsetfillcolor{currentfill}%
\pgfsetlinewidth{0.501875pt}%
\definecolor{currentstroke}{rgb}{0.000000,0.000000,0.000000}%
\pgfsetstrokecolor{currentstroke}%
\pgfsetdash{}{0pt}%
\pgfsys@defobject{currentmarker}{\pgfqpoint{0.000000in}{0.000000in}}{\pgfqpoint{0.000000in}{0.041667in}}{%
\pgfpathmoveto{\pgfqpoint{0.000000in}{0.000000in}}%
\pgfpathlineto{\pgfqpoint{0.000000in}{0.041667in}}%
\pgfusepath{stroke,fill}%
}%
\begin{pgfscope}%
\pgfsys@transformshift{2.425601in}{0.279167in}%
\pgfsys@useobject{currentmarker}{}%
\end{pgfscope}%
\end{pgfscope}%
\begin{pgfscope}%
\pgfsetbuttcap%
\pgfsetroundjoin%
\definecolor{currentfill}{rgb}{0.000000,0.000000,0.000000}%
\pgfsetfillcolor{currentfill}%
\pgfsetlinewidth{0.501875pt}%
\definecolor{currentstroke}{rgb}{0.000000,0.000000,0.000000}%
\pgfsetstrokecolor{currentstroke}%
\pgfsetdash{}{0pt}%
\pgfsys@defobject{currentmarker}{\pgfqpoint{0.000000in}{-0.041667in}}{\pgfqpoint{0.000000in}{0.000000in}}{%
\pgfpathmoveto{\pgfqpoint{0.000000in}{0.000000in}}%
\pgfpathlineto{\pgfqpoint{0.000000in}{-0.041667in}}%
\pgfusepath{stroke,fill}%
}%
\begin{pgfscope}%
\pgfsys@transformshift{2.425601in}{2.980312in}%
\pgfsys@useobject{currentmarker}{}%
\end{pgfscope}%
\end{pgfscope}%
\begin{pgfscope}%
\definecolor{textcolor}{rgb}{0.000000,0.000000,0.000000}%
\pgfsetstrokecolor{textcolor}%
\pgfsetfillcolor{textcolor}%
\pgftext[x=2.425601in,y=0.230556in,,top]{\color{textcolor}{\rmfamily\fontsize{13.000000}{15.600000}\selectfont\catcode`\^=\active\def^{\ifmmode\sp\else\^{}\fi}\catcode`\%=\active\def
\end{pgfscope}%
\begin{pgfscope}%
\pgfsetbuttcap%
\pgfsetroundjoin%
\definecolor{currentfill}{rgb}{0.000000,0.000000,0.000000}%
\pgfsetfillcolor{currentfill}%
\pgfsetlinewidth{0.501875pt}%
\definecolor{currentstroke}{rgb}{0.000000,0.000000,0.000000}%
\pgfsetstrokecolor{currentstroke}%
\pgfsetdash{}{0pt}%
\pgfsys@defobject{currentmarker}{\pgfqpoint{0.000000in}{0.000000in}}{\pgfqpoint{0.000000in}{0.041667in}}{%
\pgfpathmoveto{\pgfqpoint{0.000000in}{0.000000in}}%
\pgfpathlineto{\pgfqpoint{0.000000in}{0.041667in}}%
\pgfusepath{stroke,fill}%
}%
\begin{pgfscope}%
\pgfsys@transformshift{3.558962in}{0.279167in}%
\pgfsys@useobject{currentmarker}{}%
\end{pgfscope}%
\end{pgfscope}%
\begin{pgfscope}%
\pgfsetbuttcap%
\pgfsetroundjoin%
\definecolor{currentfill}{rgb}{0.000000,0.000000,0.000000}%
\pgfsetfillcolor{currentfill}%
\pgfsetlinewidth{0.501875pt}%
\definecolor{currentstroke}{rgb}{0.000000,0.000000,0.000000}%
\pgfsetstrokecolor{currentstroke}%
\pgfsetdash{}{0pt}%
\pgfsys@defobject{currentmarker}{\pgfqpoint{0.000000in}{-0.041667in}}{\pgfqpoint{0.000000in}{0.000000in}}{%
\pgfpathmoveto{\pgfqpoint{0.000000in}{0.000000in}}%
\pgfpathlineto{\pgfqpoint{0.000000in}{-0.041667in}}%
\pgfusepath{stroke,fill}%
}%
\begin{pgfscope}%
\pgfsys@transformshift{3.558962in}{2.980312in}%
\pgfsys@useobject{currentmarker}{}%
\end{pgfscope}%
\end{pgfscope}%
\begin{pgfscope}%
\definecolor{textcolor}{rgb}{0.000000,0.000000,0.000000}%
\pgfsetstrokecolor{textcolor}%
\pgfsetfillcolor{textcolor}%
\pgftext[x=3.558962in,y=0.230556in,,top]{\color{textcolor}{\rmfamily\fontsize{13.000000}{15.600000}\selectfont\catcode`\^=\active\def^{\ifmmode\sp\else\^{}\fi}\catcode`\%=\active\def
\end{pgfscope}%
\begin{pgfscope}%
\pgfsetbuttcap%
\pgfsetroundjoin%
\definecolor{currentfill}{rgb}{0.000000,0.000000,0.000000}%
\pgfsetfillcolor{currentfill}%
\pgfsetlinewidth{0.501875pt}%
\definecolor{currentstroke}{rgb}{0.000000,0.000000,0.000000}%
\pgfsetstrokecolor{currentstroke}%
\pgfsetdash{}{0pt}%
\pgfsys@defobject{currentmarker}{\pgfqpoint{0.000000in}{0.000000in}}{\pgfqpoint{0.000000in}{0.020833in}}{%
\pgfpathmoveto{\pgfqpoint{0.000000in}{0.000000in}}%
\pgfpathlineto{\pgfqpoint{0.000000in}{0.020833in}}%
\pgfusepath{stroke,fill}%
}%
\begin{pgfscope}%
\pgfsys@transformshift{0.838895in}{0.279167in}%
\pgfsys@useobject{currentmarker}{}%
\end{pgfscope}%
\end{pgfscope}%
\begin{pgfscope}%
\pgfsetbuttcap%
\pgfsetroundjoin%
\definecolor{currentfill}{rgb}{0.000000,0.000000,0.000000}%
\pgfsetfillcolor{currentfill}%
\pgfsetlinewidth{0.501875pt}%
\definecolor{currentstroke}{rgb}{0.000000,0.000000,0.000000}%
\pgfsetstrokecolor{currentstroke}%
\pgfsetdash{}{0pt}%
\pgfsys@defobject{currentmarker}{\pgfqpoint{0.000000in}{-0.020833in}}{\pgfqpoint{0.000000in}{0.000000in}}{%
\pgfpathmoveto{\pgfqpoint{0.000000in}{0.000000in}}%
\pgfpathlineto{\pgfqpoint{0.000000in}{-0.020833in}}%
\pgfusepath{stroke,fill}%
}%
\begin{pgfscope}%
\pgfsys@transformshift{0.838895in}{2.980312in}%
\pgfsys@useobject{currentmarker}{}%
\end{pgfscope}%
\end{pgfscope}%
\begin{pgfscope}%
\pgfsetbuttcap%
\pgfsetroundjoin%
\definecolor{currentfill}{rgb}{0.000000,0.000000,0.000000}%
\pgfsetfillcolor{currentfill}%
\pgfsetlinewidth{0.501875pt}%
\definecolor{currentstroke}{rgb}{0.000000,0.000000,0.000000}%
\pgfsetstrokecolor{currentstroke}%
\pgfsetdash{}{0pt}%
\pgfsys@defobject{currentmarker}{\pgfqpoint{0.000000in}{0.000000in}}{\pgfqpoint{0.000000in}{0.020833in}}{%
\pgfpathmoveto{\pgfqpoint{0.000000in}{0.000000in}}%
\pgfpathlineto{\pgfqpoint{0.000000in}{0.020833in}}%
\pgfusepath{stroke,fill}%
}%
\begin{pgfscope}%
\pgfsys@transformshift{1.065567in}{0.279167in}%
\pgfsys@useobject{currentmarker}{}%
\end{pgfscope}%
\end{pgfscope}%
\begin{pgfscope}%
\pgfsetbuttcap%
\pgfsetroundjoin%
\definecolor{currentfill}{rgb}{0.000000,0.000000,0.000000}%
\pgfsetfillcolor{currentfill}%
\pgfsetlinewidth{0.501875pt}%
\definecolor{currentstroke}{rgb}{0.000000,0.000000,0.000000}%
\pgfsetstrokecolor{currentstroke}%
\pgfsetdash{}{0pt}%
\pgfsys@defobject{currentmarker}{\pgfqpoint{0.000000in}{-0.020833in}}{\pgfqpoint{0.000000in}{0.000000in}}{%
\pgfpathmoveto{\pgfqpoint{0.000000in}{0.000000in}}%
\pgfpathlineto{\pgfqpoint{0.000000in}{-0.020833in}}%
\pgfusepath{stroke,fill}%
}%
\begin{pgfscope}%
\pgfsys@transformshift{1.065567in}{2.980312in}%
\pgfsys@useobject{currentmarker}{}%
\end{pgfscope}%
\end{pgfscope}%
\begin{pgfscope}%
\pgfsetbuttcap%
\pgfsetroundjoin%
\definecolor{currentfill}{rgb}{0.000000,0.000000,0.000000}%
\pgfsetfillcolor{currentfill}%
\pgfsetlinewidth{0.501875pt}%
\definecolor{currentstroke}{rgb}{0.000000,0.000000,0.000000}%
\pgfsetstrokecolor{currentstroke}%
\pgfsetdash{}{0pt}%
\pgfsys@defobject{currentmarker}{\pgfqpoint{0.000000in}{0.000000in}}{\pgfqpoint{0.000000in}{0.020833in}}{%
\pgfpathmoveto{\pgfqpoint{0.000000in}{0.000000in}}%
\pgfpathlineto{\pgfqpoint{0.000000in}{0.020833in}}%
\pgfusepath{stroke,fill}%
}%
\begin{pgfscope}%
\pgfsys@transformshift{1.518912in}{0.279167in}%
\pgfsys@useobject{currentmarker}{}%
\end{pgfscope}%
\end{pgfscope}%
\begin{pgfscope}%
\pgfsetbuttcap%
\pgfsetroundjoin%
\definecolor{currentfill}{rgb}{0.000000,0.000000,0.000000}%
\pgfsetfillcolor{currentfill}%
\pgfsetlinewidth{0.501875pt}%
\definecolor{currentstroke}{rgb}{0.000000,0.000000,0.000000}%
\pgfsetstrokecolor{currentstroke}%
\pgfsetdash{}{0pt}%
\pgfsys@defobject{currentmarker}{\pgfqpoint{0.000000in}{-0.020833in}}{\pgfqpoint{0.000000in}{0.000000in}}{%
\pgfpathmoveto{\pgfqpoint{0.000000in}{0.000000in}}%
\pgfpathlineto{\pgfqpoint{0.000000in}{-0.020833in}}%
\pgfusepath{stroke,fill}%
}%
\begin{pgfscope}%
\pgfsys@transformshift{1.518912in}{2.980312in}%
\pgfsys@useobject{currentmarker}{}%
\end{pgfscope}%
\end{pgfscope}%
\begin{pgfscope}%
\pgfsetbuttcap%
\pgfsetroundjoin%
\definecolor{currentfill}{rgb}{0.000000,0.000000,0.000000}%
\pgfsetfillcolor{currentfill}%
\pgfsetlinewidth{0.501875pt}%
\definecolor{currentstroke}{rgb}{0.000000,0.000000,0.000000}%
\pgfsetstrokecolor{currentstroke}%
\pgfsetdash{}{0pt}%
\pgfsys@defobject{currentmarker}{\pgfqpoint{0.000000in}{0.000000in}}{\pgfqpoint{0.000000in}{0.020833in}}{%
\pgfpathmoveto{\pgfqpoint{0.000000in}{0.000000in}}%
\pgfpathlineto{\pgfqpoint{0.000000in}{0.020833in}}%
\pgfusepath{stroke,fill}%
}%
\begin{pgfscope}%
\pgfsys@transformshift{1.745584in}{0.279167in}%
\pgfsys@useobject{currentmarker}{}%
\end{pgfscope}%
\end{pgfscope}%
\begin{pgfscope}%
\pgfsetbuttcap%
\pgfsetroundjoin%
\definecolor{currentfill}{rgb}{0.000000,0.000000,0.000000}%
\pgfsetfillcolor{currentfill}%
\pgfsetlinewidth{0.501875pt}%
\definecolor{currentstroke}{rgb}{0.000000,0.000000,0.000000}%
\pgfsetstrokecolor{currentstroke}%
\pgfsetdash{}{0pt}%
\pgfsys@defobject{currentmarker}{\pgfqpoint{0.000000in}{-0.020833in}}{\pgfqpoint{0.000000in}{0.000000in}}{%
\pgfpathmoveto{\pgfqpoint{0.000000in}{0.000000in}}%
\pgfpathlineto{\pgfqpoint{0.000000in}{-0.020833in}}%
\pgfusepath{stroke,fill}%
}%
\begin{pgfscope}%
\pgfsys@transformshift{1.745584in}{2.980312in}%
\pgfsys@useobject{currentmarker}{}%
\end{pgfscope}%
\end{pgfscope}%
\begin{pgfscope}%
\pgfsetbuttcap%
\pgfsetroundjoin%
\definecolor{currentfill}{rgb}{0.000000,0.000000,0.000000}%
\pgfsetfillcolor{currentfill}%
\pgfsetlinewidth{0.501875pt}%
\definecolor{currentstroke}{rgb}{0.000000,0.000000,0.000000}%
\pgfsetstrokecolor{currentstroke}%
\pgfsetdash{}{0pt}%
\pgfsys@defobject{currentmarker}{\pgfqpoint{0.000000in}{0.000000in}}{\pgfqpoint{0.000000in}{0.020833in}}{%
\pgfpathmoveto{\pgfqpoint{0.000000in}{0.000000in}}%
\pgfpathlineto{\pgfqpoint{0.000000in}{0.020833in}}%
\pgfusepath{stroke,fill}%
}%
\begin{pgfscope}%
\pgfsys@transformshift{1.972256in}{0.279167in}%
\pgfsys@useobject{currentmarker}{}%
\end{pgfscope}%
\end{pgfscope}%
\begin{pgfscope}%
\pgfsetbuttcap%
\pgfsetroundjoin%
\definecolor{currentfill}{rgb}{0.000000,0.000000,0.000000}%
\pgfsetfillcolor{currentfill}%
\pgfsetlinewidth{0.501875pt}%
\definecolor{currentstroke}{rgb}{0.000000,0.000000,0.000000}%
\pgfsetstrokecolor{currentstroke}%
\pgfsetdash{}{0pt}%
\pgfsys@defobject{currentmarker}{\pgfqpoint{0.000000in}{-0.020833in}}{\pgfqpoint{0.000000in}{0.000000in}}{%
\pgfpathmoveto{\pgfqpoint{0.000000in}{0.000000in}}%
\pgfpathlineto{\pgfqpoint{0.000000in}{-0.020833in}}%
\pgfusepath{stroke,fill}%
}%
\begin{pgfscope}%
\pgfsys@transformshift{1.972256in}{2.980312in}%
\pgfsys@useobject{currentmarker}{}%
\end{pgfscope}%
\end{pgfscope}%
\begin{pgfscope}%
\pgfsetbuttcap%
\pgfsetroundjoin%
\definecolor{currentfill}{rgb}{0.000000,0.000000,0.000000}%
\pgfsetfillcolor{currentfill}%
\pgfsetlinewidth{0.501875pt}%
\definecolor{currentstroke}{rgb}{0.000000,0.000000,0.000000}%
\pgfsetstrokecolor{currentstroke}%
\pgfsetdash{}{0pt}%
\pgfsys@defobject{currentmarker}{\pgfqpoint{0.000000in}{0.000000in}}{\pgfqpoint{0.000000in}{0.020833in}}{%
\pgfpathmoveto{\pgfqpoint{0.000000in}{0.000000in}}%
\pgfpathlineto{\pgfqpoint{0.000000in}{0.020833in}}%
\pgfusepath{stroke,fill}%
}%
\begin{pgfscope}%
\pgfsys@transformshift{2.198929in}{0.279167in}%
\pgfsys@useobject{currentmarker}{}%
\end{pgfscope}%
\end{pgfscope}%
\begin{pgfscope}%
\pgfsetbuttcap%
\pgfsetroundjoin%
\definecolor{currentfill}{rgb}{0.000000,0.000000,0.000000}%
\pgfsetfillcolor{currentfill}%
\pgfsetlinewidth{0.501875pt}%
\definecolor{currentstroke}{rgb}{0.000000,0.000000,0.000000}%
\pgfsetstrokecolor{currentstroke}%
\pgfsetdash{}{0pt}%
\pgfsys@defobject{currentmarker}{\pgfqpoint{0.000000in}{-0.020833in}}{\pgfqpoint{0.000000in}{0.000000in}}{%
\pgfpathmoveto{\pgfqpoint{0.000000in}{0.000000in}}%
\pgfpathlineto{\pgfqpoint{0.000000in}{-0.020833in}}%
\pgfusepath{stroke,fill}%
}%
\begin{pgfscope}%
\pgfsys@transformshift{2.198929in}{2.980312in}%
\pgfsys@useobject{currentmarker}{}%
\end{pgfscope}%
\end{pgfscope}%
\begin{pgfscope}%
\pgfsetbuttcap%
\pgfsetroundjoin%
\definecolor{currentfill}{rgb}{0.000000,0.000000,0.000000}%
\pgfsetfillcolor{currentfill}%
\pgfsetlinewidth{0.501875pt}%
\definecolor{currentstroke}{rgb}{0.000000,0.000000,0.000000}%
\pgfsetstrokecolor{currentstroke}%
\pgfsetdash{}{0pt}%
\pgfsys@defobject{currentmarker}{\pgfqpoint{0.000000in}{0.000000in}}{\pgfqpoint{0.000000in}{0.020833in}}{%
\pgfpathmoveto{\pgfqpoint{0.000000in}{0.000000in}}%
\pgfpathlineto{\pgfqpoint{0.000000in}{0.020833in}}%
\pgfusepath{stroke,fill}%
}%
\begin{pgfscope}%
\pgfsys@transformshift{2.652273in}{0.279167in}%
\pgfsys@useobject{currentmarker}{}%
\end{pgfscope}%
\end{pgfscope}%
\begin{pgfscope}%
\pgfsetbuttcap%
\pgfsetroundjoin%
\definecolor{currentfill}{rgb}{0.000000,0.000000,0.000000}%
\pgfsetfillcolor{currentfill}%
\pgfsetlinewidth{0.501875pt}%
\definecolor{currentstroke}{rgb}{0.000000,0.000000,0.000000}%
\pgfsetstrokecolor{currentstroke}%
\pgfsetdash{}{0pt}%
\pgfsys@defobject{currentmarker}{\pgfqpoint{0.000000in}{-0.020833in}}{\pgfqpoint{0.000000in}{0.000000in}}{%
\pgfpathmoveto{\pgfqpoint{0.000000in}{0.000000in}}%
\pgfpathlineto{\pgfqpoint{0.000000in}{-0.020833in}}%
\pgfusepath{stroke,fill}%
}%
\begin{pgfscope}%
\pgfsys@transformshift{2.652273in}{2.980312in}%
\pgfsys@useobject{currentmarker}{}%
\end{pgfscope}%
\end{pgfscope}%
\begin{pgfscope}%
\pgfsetbuttcap%
\pgfsetroundjoin%
\definecolor{currentfill}{rgb}{0.000000,0.000000,0.000000}%
\pgfsetfillcolor{currentfill}%
\pgfsetlinewidth{0.501875pt}%
\definecolor{currentstroke}{rgb}{0.000000,0.000000,0.000000}%
\pgfsetstrokecolor{currentstroke}%
\pgfsetdash{}{0pt}%
\pgfsys@defobject{currentmarker}{\pgfqpoint{0.000000in}{0.000000in}}{\pgfqpoint{0.000000in}{0.020833in}}{%
\pgfpathmoveto{\pgfqpoint{0.000000in}{0.000000in}}%
\pgfpathlineto{\pgfqpoint{0.000000in}{0.020833in}}%
\pgfusepath{stroke,fill}%
}%
\begin{pgfscope}%
\pgfsys@transformshift{2.878945in}{0.279167in}%
\pgfsys@useobject{currentmarker}{}%
\end{pgfscope}%
\end{pgfscope}%
\begin{pgfscope}%
\pgfsetbuttcap%
\pgfsetroundjoin%
\definecolor{currentfill}{rgb}{0.000000,0.000000,0.000000}%
\pgfsetfillcolor{currentfill}%
\pgfsetlinewidth{0.501875pt}%
\definecolor{currentstroke}{rgb}{0.000000,0.000000,0.000000}%
\pgfsetstrokecolor{currentstroke}%
\pgfsetdash{}{0pt}%
\pgfsys@defobject{currentmarker}{\pgfqpoint{0.000000in}{-0.020833in}}{\pgfqpoint{0.000000in}{0.000000in}}{%
\pgfpathmoveto{\pgfqpoint{0.000000in}{0.000000in}}%
\pgfpathlineto{\pgfqpoint{0.000000in}{-0.020833in}}%
\pgfusepath{stroke,fill}%
}%
\begin{pgfscope}%
\pgfsys@transformshift{2.878945in}{2.980312in}%
\pgfsys@useobject{currentmarker}{}%
\end{pgfscope}%
\end{pgfscope}%
\begin{pgfscope}%
\pgfsetbuttcap%
\pgfsetroundjoin%
\definecolor{currentfill}{rgb}{0.000000,0.000000,0.000000}%
\pgfsetfillcolor{currentfill}%
\pgfsetlinewidth{0.501875pt}%
\definecolor{currentstroke}{rgb}{0.000000,0.000000,0.000000}%
\pgfsetstrokecolor{currentstroke}%
\pgfsetdash{}{0pt}%
\pgfsys@defobject{currentmarker}{\pgfqpoint{0.000000in}{0.000000in}}{\pgfqpoint{0.000000in}{0.020833in}}{%
\pgfpathmoveto{\pgfqpoint{0.000000in}{0.000000in}}%
\pgfpathlineto{\pgfqpoint{0.000000in}{0.020833in}}%
\pgfusepath{stroke,fill}%
}%
\begin{pgfscope}%
\pgfsys@transformshift{3.105618in}{0.279167in}%
\pgfsys@useobject{currentmarker}{}%
\end{pgfscope}%
\end{pgfscope}%
\begin{pgfscope}%
\pgfsetbuttcap%
\pgfsetroundjoin%
\definecolor{currentfill}{rgb}{0.000000,0.000000,0.000000}%
\pgfsetfillcolor{currentfill}%
\pgfsetlinewidth{0.501875pt}%
\definecolor{currentstroke}{rgb}{0.000000,0.000000,0.000000}%
\pgfsetstrokecolor{currentstroke}%
\pgfsetdash{}{0pt}%
\pgfsys@defobject{currentmarker}{\pgfqpoint{0.000000in}{-0.020833in}}{\pgfqpoint{0.000000in}{0.000000in}}{%
\pgfpathmoveto{\pgfqpoint{0.000000in}{0.000000in}}%
\pgfpathlineto{\pgfqpoint{0.000000in}{-0.020833in}}%
\pgfusepath{stroke,fill}%
}%
\begin{pgfscope}%
\pgfsys@transformshift{3.105618in}{2.980312in}%
\pgfsys@useobject{currentmarker}{}%
\end{pgfscope}%
\end{pgfscope}%
\begin{pgfscope}%
\pgfsetbuttcap%
\pgfsetroundjoin%
\definecolor{currentfill}{rgb}{0.000000,0.000000,0.000000}%
\pgfsetfillcolor{currentfill}%
\pgfsetlinewidth{0.501875pt}%
\definecolor{currentstroke}{rgb}{0.000000,0.000000,0.000000}%
\pgfsetstrokecolor{currentstroke}%
\pgfsetdash{}{0pt}%
\pgfsys@defobject{currentmarker}{\pgfqpoint{0.000000in}{0.000000in}}{\pgfqpoint{0.000000in}{0.020833in}}{%
\pgfpathmoveto{\pgfqpoint{0.000000in}{0.000000in}}%
\pgfpathlineto{\pgfqpoint{0.000000in}{0.020833in}}%
\pgfusepath{stroke,fill}%
}%
\begin{pgfscope}%
\pgfsys@transformshift{3.332290in}{0.279167in}%
\pgfsys@useobject{currentmarker}{}%
\end{pgfscope}%
\end{pgfscope}%
\begin{pgfscope}%
\pgfsetbuttcap%
\pgfsetroundjoin%
\definecolor{currentfill}{rgb}{0.000000,0.000000,0.000000}%
\pgfsetfillcolor{currentfill}%
\pgfsetlinewidth{0.501875pt}%
\definecolor{currentstroke}{rgb}{0.000000,0.000000,0.000000}%
\pgfsetstrokecolor{currentstroke}%
\pgfsetdash{}{0pt}%
\pgfsys@defobject{currentmarker}{\pgfqpoint{0.000000in}{-0.020833in}}{\pgfqpoint{0.000000in}{0.000000in}}{%
\pgfpathmoveto{\pgfqpoint{0.000000in}{0.000000in}}%
\pgfpathlineto{\pgfqpoint{0.000000in}{-0.020833in}}%
\pgfusepath{stroke,fill}%
}%
\begin{pgfscope}%
\pgfsys@transformshift{3.332290in}{2.980312in}%
\pgfsys@useobject{currentmarker}{}%
\end{pgfscope}%
\end{pgfscope}%
\begin{pgfscope}%
\pgfsetbuttcap%
\pgfsetroundjoin%
\definecolor{currentfill}{rgb}{0.000000,0.000000,0.000000}%
\pgfsetfillcolor{currentfill}%
\pgfsetlinewidth{0.501875pt}%
\definecolor{currentstroke}{rgb}{0.000000,0.000000,0.000000}%
\pgfsetstrokecolor{currentstroke}%
\pgfsetdash{}{0pt}%
\pgfsys@defobject{currentmarker}{\pgfqpoint{0.000000in}{0.000000in}}{\pgfqpoint{0.000000in}{0.020833in}}{%
\pgfpathmoveto{\pgfqpoint{0.000000in}{0.000000in}}%
\pgfpathlineto{\pgfqpoint{0.000000in}{0.020833in}}%
\pgfusepath{stroke,fill}%
}%
\begin{pgfscope}%
\pgfsys@transformshift{3.785635in}{0.279167in}%
\pgfsys@useobject{currentmarker}{}%
\end{pgfscope}%
\end{pgfscope}%
\begin{pgfscope}%
\pgfsetbuttcap%
\pgfsetroundjoin%
\definecolor{currentfill}{rgb}{0.000000,0.000000,0.000000}%
\pgfsetfillcolor{currentfill}%
\pgfsetlinewidth{0.501875pt}%
\definecolor{currentstroke}{rgb}{0.000000,0.000000,0.000000}%
\pgfsetstrokecolor{currentstroke}%
\pgfsetdash{}{0pt}%
\pgfsys@defobject{currentmarker}{\pgfqpoint{0.000000in}{-0.020833in}}{\pgfqpoint{0.000000in}{0.000000in}}{%
\pgfpathmoveto{\pgfqpoint{0.000000in}{0.000000in}}%
\pgfpathlineto{\pgfqpoint{0.000000in}{-0.020833in}}%
\pgfusepath{stroke,fill}%
}%
\begin{pgfscope}%
\pgfsys@transformshift{3.785635in}{2.980312in}%
\pgfsys@useobject{currentmarker}{}%
\end{pgfscope}%
\end{pgfscope}%
\begin{pgfscope}%
\pgfsetbuttcap%
\pgfsetroundjoin%
\definecolor{currentfill}{rgb}{0.000000,0.000000,0.000000}%
\pgfsetfillcolor{currentfill}%
\pgfsetlinewidth{0.501875pt}%
\definecolor{currentstroke}{rgb}{0.000000,0.000000,0.000000}%
\pgfsetstrokecolor{currentstroke}%
\pgfsetdash{}{0pt}%
\pgfsys@defobject{currentmarker}{\pgfqpoint{0.000000in}{0.000000in}}{\pgfqpoint{0.000000in}{0.020833in}}{%
\pgfpathmoveto{\pgfqpoint{0.000000in}{0.000000in}}%
\pgfpathlineto{\pgfqpoint{0.000000in}{0.020833in}}%
\pgfusepath{stroke,fill}%
}%
\begin{pgfscope}%
\pgfsys@transformshift{4.012307in}{0.279167in}%
\pgfsys@useobject{currentmarker}{}%
\end{pgfscope}%
\end{pgfscope}%
\begin{pgfscope}%
\pgfsetbuttcap%
\pgfsetroundjoin%
\definecolor{currentfill}{rgb}{0.000000,0.000000,0.000000}%
\pgfsetfillcolor{currentfill}%
\pgfsetlinewidth{0.501875pt}%
\definecolor{currentstroke}{rgb}{0.000000,0.000000,0.000000}%
\pgfsetstrokecolor{currentstroke}%
\pgfsetdash{}{0pt}%
\pgfsys@defobject{currentmarker}{\pgfqpoint{0.000000in}{-0.020833in}}{\pgfqpoint{0.000000in}{0.000000in}}{%
\pgfpathmoveto{\pgfqpoint{0.000000in}{0.000000in}}%
\pgfpathlineto{\pgfqpoint{0.000000in}{-0.020833in}}%
\pgfusepath{stroke,fill}%
}%
\begin{pgfscope}%
\pgfsys@transformshift{4.012307in}{2.980312in}%
\pgfsys@useobject{currentmarker}{}%
\end{pgfscope}%
\end{pgfscope}%
\begin{pgfscope}%
\pgfsetbuttcap%
\pgfsetroundjoin%
\definecolor{currentfill}{rgb}{0.000000,0.000000,0.000000}%
\pgfsetfillcolor{currentfill}%
\pgfsetlinewidth{0.501875pt}%
\definecolor{currentstroke}{rgb}{0.000000,0.000000,0.000000}%
\pgfsetstrokecolor{currentstroke}%
\pgfsetdash{}{0pt}%
\pgfsys@defobject{currentmarker}{\pgfqpoint{0.000000in}{0.000000in}}{\pgfqpoint{0.041667in}{0.000000in}}{%
\pgfpathmoveto{\pgfqpoint{0.000000in}{0.000000in}}%
\pgfpathlineto{\pgfqpoint{0.041667in}{0.000000in}}%
\pgfusepath{stroke,fill}%
}%
\begin{pgfscope}%
\pgfsys@transformshift{0.773727in}{2.212329in}%
\pgfsys@useobject{currentmarker}{}%
\end{pgfscope}%
\end{pgfscope}%
\begin{pgfscope}%
\pgfsetbuttcap%
\pgfsetroundjoin%
\definecolor{currentfill}{rgb}{0.000000,0.000000,0.000000}%
\pgfsetfillcolor{currentfill}%
\pgfsetlinewidth{0.501875pt}%
\definecolor{currentstroke}{rgb}{0.000000,0.000000,0.000000}%
\pgfsetstrokecolor{currentstroke}%
\pgfsetdash{}{0pt}%
\pgfsys@defobject{currentmarker}{\pgfqpoint{-0.041667in}{0.000000in}}{\pgfqpoint{-0.000000in}{0.000000in}}{%
\pgfpathmoveto{\pgfqpoint{-0.000000in}{0.000000in}}%
\pgfpathlineto{\pgfqpoint{-0.041667in}{0.000000in}}%
\pgfusepath{stroke,fill}%
}%
\begin{pgfscope}%
\pgfsys@transformshift{4.077475in}{2.212329in}%
\pgfsys@useobject{currentmarker}{}%
\end{pgfscope}%
\end{pgfscope}%
\begin{pgfscope}%
\definecolor{textcolor}{rgb}{0.000000,0.000000,0.000000}%
\pgfsetstrokecolor{textcolor}%
\pgfsetfillcolor{textcolor}%
\pgftext[x=0.523919in, y=2.152329in, left, base]{\color{textcolor}{\rmfamily\fontsize{10.000000}{12.000000}\selectfont\catcode`\^=\active\def^{\ifmmode\sp\else\^{}\fi}\catcode`\%=\active\def
\end{pgfscope}%
\begin{pgfscope}%
\pgfsetbuttcap%
\pgfsetroundjoin%
\definecolor{currentfill}{rgb}{0.000000,0.000000,0.000000}%
\pgfsetfillcolor{currentfill}%
\pgfsetlinewidth{0.501875pt}%
\definecolor{currentstroke}{rgb}{0.000000,0.000000,0.000000}%
\pgfsetstrokecolor{currentstroke}%
\pgfsetdash{}{0pt}%
\pgfsys@defobject{currentmarker}{\pgfqpoint{0.000000in}{0.000000in}}{\pgfqpoint{0.020833in}{0.000000in}}{%
\pgfpathmoveto{\pgfqpoint{0.000000in}{0.000000in}}%
\pgfpathlineto{\pgfqpoint{0.020833in}{0.000000in}}%
\pgfusepath{stroke,fill}%
}%
\begin{pgfscope}%
\pgfsys@transformshift{0.773727in}{0.829958in}%
\pgfsys@useobject{currentmarker}{}%
\end{pgfscope}%
\end{pgfscope}%
\begin{pgfscope}%
\pgfsetbuttcap%
\pgfsetroundjoin%
\definecolor{currentfill}{rgb}{0.000000,0.000000,0.000000}%
\pgfsetfillcolor{currentfill}%
\pgfsetlinewidth{0.501875pt}%
\definecolor{currentstroke}{rgb}{0.000000,0.000000,0.000000}%
\pgfsetstrokecolor{currentstroke}%
\pgfsetdash{}{0pt}%
\pgfsys@defobject{currentmarker}{\pgfqpoint{-0.020833in}{0.000000in}}{\pgfqpoint{-0.000000in}{0.000000in}}{%
\pgfpathmoveto{\pgfqpoint{-0.000000in}{0.000000in}}%
\pgfpathlineto{\pgfqpoint{-0.020833in}{0.000000in}}%
\pgfusepath{stroke,fill}%
}%
\begin{pgfscope}%
\pgfsys@transformshift{4.077475in}{0.829958in}%
\pgfsys@useobject{currentmarker}{}%
\end{pgfscope}%
\end{pgfscope}%
\begin{pgfscope}%
\definecolor{textcolor}{rgb}{0.000000,0.000000,0.000000}%
\pgfsetstrokecolor{textcolor}%
\pgfsetfillcolor{textcolor}%
\pgftext[x=0.286111in, y=0.769958in, left, base]{\color{textcolor}{\rmfamily\fontsize{10.000000}{12.000000}\selectfont\catcode`\^=\active\def^{\ifmmode\sp\else\^{}\fi}\catcode`\%=\active\def
\end{pgfscope}%
\begin{pgfscope}%
\pgfsetbuttcap%
\pgfsetroundjoin%
\definecolor{currentfill}{rgb}{0.000000,0.000000,0.000000}%
\pgfsetfillcolor{currentfill}%
\pgfsetlinewidth{0.501875pt}%
\definecolor{currentstroke}{rgb}{0.000000,0.000000,0.000000}%
\pgfsetstrokecolor{currentstroke}%
\pgfsetdash{}{0pt}%
\pgfsys@defobject{currentmarker}{\pgfqpoint{0.000000in}{0.000000in}}{\pgfqpoint{0.020833in}{0.000000in}}{%
\pgfpathmoveto{\pgfqpoint{0.000000in}{0.000000in}}%
\pgfpathlineto{\pgfqpoint{0.020833in}{0.000000in}}%
\pgfusepath{stroke,fill}%
}%
\begin{pgfscope}%
\pgfsys@transformshift{0.773727in}{1.178218in}%
\pgfsys@useobject{currentmarker}{}%
\end{pgfscope}%
\end{pgfscope}%
\begin{pgfscope}%
\pgfsetbuttcap%
\pgfsetroundjoin%
\definecolor{currentfill}{rgb}{0.000000,0.000000,0.000000}%
\pgfsetfillcolor{currentfill}%
\pgfsetlinewidth{0.501875pt}%
\definecolor{currentstroke}{rgb}{0.000000,0.000000,0.000000}%
\pgfsetstrokecolor{currentstroke}%
\pgfsetdash{}{0pt}%
\pgfsys@defobject{currentmarker}{\pgfqpoint{-0.020833in}{0.000000in}}{\pgfqpoint{-0.000000in}{0.000000in}}{%
\pgfpathmoveto{\pgfqpoint{-0.000000in}{0.000000in}}%
\pgfpathlineto{\pgfqpoint{-0.020833in}{0.000000in}}%
\pgfusepath{stroke,fill}%
}%
\begin{pgfscope}%
\pgfsys@transformshift{4.077475in}{1.178218in}%
\pgfsys@useobject{currentmarker}{}%
\end{pgfscope}%
\end{pgfscope}%
\begin{pgfscope}%
\definecolor{textcolor}{rgb}{0.000000,0.000000,0.000000}%
\pgfsetstrokecolor{textcolor}%
\pgfsetfillcolor{textcolor}%
\pgftext[x=0.286111in, y=1.118218in, left, base]{\color{textcolor}{\rmfamily\fontsize{10.000000}{12.000000}\selectfont\catcode`\^=\active\def^{\ifmmode\sp\else\^{}\fi}\catcode`\%=\active\def
\end{pgfscope}%
\begin{pgfscope}%
\pgfsetbuttcap%
\pgfsetroundjoin%
\definecolor{currentfill}{rgb}{0.000000,0.000000,0.000000}%
\pgfsetfillcolor{currentfill}%
\pgfsetlinewidth{0.501875pt}%
\definecolor{currentstroke}{rgb}{0.000000,0.000000,0.000000}%
\pgfsetstrokecolor{currentstroke}%
\pgfsetdash{}{0pt}%
\pgfsys@defobject{currentmarker}{\pgfqpoint{0.000000in}{0.000000in}}{\pgfqpoint{0.020833in}{0.000000in}}{%
\pgfpathmoveto{\pgfqpoint{0.000000in}{0.000000in}}%
\pgfpathlineto{\pgfqpoint{0.020833in}{0.000000in}}%
\pgfusepath{stroke,fill}%
}%
\begin{pgfscope}%
\pgfsys@transformshift{0.773727in}{1.425313in}%
\pgfsys@useobject{currentmarker}{}%
\end{pgfscope}%
\end{pgfscope}%
\begin{pgfscope}%
\pgfsetbuttcap%
\pgfsetroundjoin%
\definecolor{currentfill}{rgb}{0.000000,0.000000,0.000000}%
\pgfsetfillcolor{currentfill}%
\pgfsetlinewidth{0.501875pt}%
\definecolor{currentstroke}{rgb}{0.000000,0.000000,0.000000}%
\pgfsetstrokecolor{currentstroke}%
\pgfsetdash{}{0pt}%
\pgfsys@defobject{currentmarker}{\pgfqpoint{-0.020833in}{0.000000in}}{\pgfqpoint{-0.000000in}{0.000000in}}{%
\pgfpathmoveto{\pgfqpoint{-0.000000in}{0.000000in}}%
\pgfpathlineto{\pgfqpoint{-0.020833in}{0.000000in}}%
\pgfusepath{stroke,fill}%
}%
\begin{pgfscope}%
\pgfsys@transformshift{4.077475in}{1.425313in}%
\pgfsys@useobject{currentmarker}{}%
\end{pgfscope}%
\end{pgfscope}%
\begin{pgfscope}%
\definecolor{textcolor}{rgb}{0.000000,0.000000,0.000000}%
\pgfsetstrokecolor{textcolor}%
\pgfsetfillcolor{textcolor}%
\pgftext[x=0.286111in, y=1.365313in, left, base]{\color{textcolor}{\rmfamily\fontsize{10.000000}{12.000000}\selectfont\catcode`\^=\active\def^{\ifmmode\sp\else\^{}\fi}\catcode`\%=\active\def
\end{pgfscope}%
\begin{pgfscope}%
\pgfsetbuttcap%
\pgfsetroundjoin%
\definecolor{currentfill}{rgb}{0.000000,0.000000,0.000000}%
\pgfsetfillcolor{currentfill}%
\pgfsetlinewidth{0.501875pt}%
\definecolor{currentstroke}{rgb}{0.000000,0.000000,0.000000}%
\pgfsetstrokecolor{currentstroke}%
\pgfsetdash{}{0pt}%
\pgfsys@defobject{currentmarker}{\pgfqpoint{0.000000in}{0.000000in}}{\pgfqpoint{0.020833in}{0.000000in}}{%
\pgfpathmoveto{\pgfqpoint{0.000000in}{0.000000in}}%
\pgfpathlineto{\pgfqpoint{0.020833in}{0.000000in}}%
\pgfusepath{stroke,fill}%
}%
\begin{pgfscope}%
\pgfsys@transformshift{0.773727in}{1.616974in}%
\pgfsys@useobject{currentmarker}{}%
\end{pgfscope}%
\end{pgfscope}%
\begin{pgfscope}%
\pgfsetbuttcap%
\pgfsetroundjoin%
\definecolor{currentfill}{rgb}{0.000000,0.000000,0.000000}%
\pgfsetfillcolor{currentfill}%
\pgfsetlinewidth{0.501875pt}%
\definecolor{currentstroke}{rgb}{0.000000,0.000000,0.000000}%
\pgfsetstrokecolor{currentstroke}%
\pgfsetdash{}{0pt}%
\pgfsys@defobject{currentmarker}{\pgfqpoint{-0.020833in}{0.000000in}}{\pgfqpoint{-0.000000in}{0.000000in}}{%
\pgfpathmoveto{\pgfqpoint{-0.000000in}{0.000000in}}%
\pgfpathlineto{\pgfqpoint{-0.020833in}{0.000000in}}%
\pgfusepath{stroke,fill}%
}%
\begin{pgfscope}%
\pgfsys@transformshift{4.077475in}{1.616974in}%
\pgfsys@useobject{currentmarker}{}%
\end{pgfscope}%
\end{pgfscope}%
\begin{pgfscope}%
\pgfsetbuttcap%
\pgfsetroundjoin%
\definecolor{currentfill}{rgb}{0.000000,0.000000,0.000000}%
\pgfsetfillcolor{currentfill}%
\pgfsetlinewidth{0.501875pt}%
\definecolor{currentstroke}{rgb}{0.000000,0.000000,0.000000}%
\pgfsetstrokecolor{currentstroke}%
\pgfsetdash{}{0pt}%
\pgfsys@defobject{currentmarker}{\pgfqpoint{0.000000in}{0.000000in}}{\pgfqpoint{0.020833in}{0.000000in}}{%
\pgfpathmoveto{\pgfqpoint{0.000000in}{0.000000in}}%
\pgfpathlineto{\pgfqpoint{0.020833in}{0.000000in}}%
\pgfusepath{stroke,fill}%
}%
\begin{pgfscope}%
\pgfsys@transformshift{0.773727in}{1.773573in}%
\pgfsys@useobject{currentmarker}{}%
\end{pgfscope}%
\end{pgfscope}%
\begin{pgfscope}%
\pgfsetbuttcap%
\pgfsetroundjoin%
\definecolor{currentfill}{rgb}{0.000000,0.000000,0.000000}%
\pgfsetfillcolor{currentfill}%
\pgfsetlinewidth{0.501875pt}%
\definecolor{currentstroke}{rgb}{0.000000,0.000000,0.000000}%
\pgfsetstrokecolor{currentstroke}%
\pgfsetdash{}{0pt}%
\pgfsys@defobject{currentmarker}{\pgfqpoint{-0.020833in}{0.000000in}}{\pgfqpoint{-0.000000in}{0.000000in}}{%
\pgfpathmoveto{\pgfqpoint{-0.000000in}{0.000000in}}%
\pgfpathlineto{\pgfqpoint{-0.020833in}{0.000000in}}%
\pgfusepath{stroke,fill}%
}%
\begin{pgfscope}%
\pgfsys@transformshift{4.077475in}{1.773573in}%
\pgfsys@useobject{currentmarker}{}%
\end{pgfscope}%
\end{pgfscope}%
\begin{pgfscope}%
\definecolor{textcolor}{rgb}{0.000000,0.000000,0.000000}%
\pgfsetstrokecolor{textcolor}%
\pgfsetfillcolor{textcolor}%
\pgftext[x=0.286111in, y=1.713573in, left, base]{\color{textcolor}{\rmfamily\fontsize{10.000000}{12.000000}\selectfont\catcode`\^=\active\def^{\ifmmode\sp\else\^{}\fi}\catcode`\%=\active\def
\end{pgfscope}%
\begin{pgfscope}%
\pgfsetbuttcap%
\pgfsetroundjoin%
\definecolor{currentfill}{rgb}{0.000000,0.000000,0.000000}%
\pgfsetfillcolor{currentfill}%
\pgfsetlinewidth{0.501875pt}%
\definecolor{currentstroke}{rgb}{0.000000,0.000000,0.000000}%
\pgfsetstrokecolor{currentstroke}%
\pgfsetdash{}{0pt}%
\pgfsys@defobject{currentmarker}{\pgfqpoint{0.000000in}{0.000000in}}{\pgfqpoint{0.020833in}{0.000000in}}{%
\pgfpathmoveto{\pgfqpoint{0.000000in}{0.000000in}}%
\pgfpathlineto{\pgfqpoint{0.020833in}{0.000000in}}%
\pgfusepath{stroke,fill}%
}%
\begin{pgfscope}%
\pgfsys@transformshift{0.773727in}{1.905976in}%
\pgfsys@useobject{currentmarker}{}%
\end{pgfscope}%
\end{pgfscope}%
\begin{pgfscope}%
\pgfsetbuttcap%
\pgfsetroundjoin%
\definecolor{currentfill}{rgb}{0.000000,0.000000,0.000000}%
\pgfsetfillcolor{currentfill}%
\pgfsetlinewidth{0.501875pt}%
\definecolor{currentstroke}{rgb}{0.000000,0.000000,0.000000}%
\pgfsetstrokecolor{currentstroke}%
\pgfsetdash{}{0pt}%
\pgfsys@defobject{currentmarker}{\pgfqpoint{-0.020833in}{0.000000in}}{\pgfqpoint{-0.000000in}{0.000000in}}{%
\pgfpathmoveto{\pgfqpoint{-0.000000in}{0.000000in}}%
\pgfpathlineto{\pgfqpoint{-0.020833in}{0.000000in}}%
\pgfusepath{stroke,fill}%
}%
\begin{pgfscope}%
\pgfsys@transformshift{4.077475in}{1.905976in}%
\pgfsys@useobject{currentmarker}{}%
\end{pgfscope}%
\end{pgfscope}%
\begin{pgfscope}%
\pgfsetbuttcap%
\pgfsetroundjoin%
\definecolor{currentfill}{rgb}{0.000000,0.000000,0.000000}%
\pgfsetfillcolor{currentfill}%
\pgfsetlinewidth{0.501875pt}%
\definecolor{currentstroke}{rgb}{0.000000,0.000000,0.000000}%
\pgfsetstrokecolor{currentstroke}%
\pgfsetdash{}{0pt}%
\pgfsys@defobject{currentmarker}{\pgfqpoint{0.000000in}{0.000000in}}{\pgfqpoint{0.020833in}{0.000000in}}{%
\pgfpathmoveto{\pgfqpoint{0.000000in}{0.000000in}}%
\pgfpathlineto{\pgfqpoint{0.020833in}{0.000000in}}%
\pgfusepath{stroke,fill}%
}%
\begin{pgfscope}%
\pgfsys@transformshift{0.773727in}{2.020668in}%
\pgfsys@useobject{currentmarker}{}%
\end{pgfscope}%
\end{pgfscope}%
\begin{pgfscope}%
\pgfsetbuttcap%
\pgfsetroundjoin%
\definecolor{currentfill}{rgb}{0.000000,0.000000,0.000000}%
\pgfsetfillcolor{currentfill}%
\pgfsetlinewidth{0.501875pt}%
\definecolor{currentstroke}{rgb}{0.000000,0.000000,0.000000}%
\pgfsetstrokecolor{currentstroke}%
\pgfsetdash{}{0pt}%
\pgfsys@defobject{currentmarker}{\pgfqpoint{-0.020833in}{0.000000in}}{\pgfqpoint{-0.000000in}{0.000000in}}{%
\pgfpathmoveto{\pgfqpoint{-0.000000in}{0.000000in}}%
\pgfpathlineto{\pgfqpoint{-0.020833in}{0.000000in}}%
\pgfusepath{stroke,fill}%
}%
\begin{pgfscope}%
\pgfsys@transformshift{4.077475in}{2.020668in}%
\pgfsys@useobject{currentmarker}{}%
\end{pgfscope}%
\end{pgfscope}%
\begin{pgfscope}%
\pgfsetbuttcap%
\pgfsetroundjoin%
\definecolor{currentfill}{rgb}{0.000000,0.000000,0.000000}%
\pgfsetfillcolor{currentfill}%
\pgfsetlinewidth{0.501875pt}%
\definecolor{currentstroke}{rgb}{0.000000,0.000000,0.000000}%
\pgfsetstrokecolor{currentstroke}%
\pgfsetdash{}{0pt}%
\pgfsys@defobject{currentmarker}{\pgfqpoint{0.000000in}{0.000000in}}{\pgfqpoint{0.020833in}{0.000000in}}{%
\pgfpathmoveto{\pgfqpoint{0.000000in}{0.000000in}}%
\pgfpathlineto{\pgfqpoint{0.020833in}{0.000000in}}%
\pgfusepath{stroke,fill}%
}%
\begin{pgfscope}%
\pgfsys@transformshift{0.773727in}{2.121834in}%
\pgfsys@useobject{currentmarker}{}%
\end{pgfscope}%
\end{pgfscope}%
\begin{pgfscope}%
\pgfsetbuttcap%
\pgfsetroundjoin%
\definecolor{currentfill}{rgb}{0.000000,0.000000,0.000000}%
\pgfsetfillcolor{currentfill}%
\pgfsetlinewidth{0.501875pt}%
\definecolor{currentstroke}{rgb}{0.000000,0.000000,0.000000}%
\pgfsetstrokecolor{currentstroke}%
\pgfsetdash{}{0pt}%
\pgfsys@defobject{currentmarker}{\pgfqpoint{-0.020833in}{0.000000in}}{\pgfqpoint{-0.000000in}{0.000000in}}{%
\pgfpathmoveto{\pgfqpoint{-0.000000in}{0.000000in}}%
\pgfpathlineto{\pgfqpoint{-0.020833in}{0.000000in}}%
\pgfusepath{stroke,fill}%
}%
\begin{pgfscope}%
\pgfsys@transformshift{4.077475in}{2.121834in}%
\pgfsys@useobject{currentmarker}{}%
\end{pgfscope}%
\end{pgfscope}%
\begin{pgfscope}%
\pgfsetbuttcap%
\pgfsetroundjoin%
\definecolor{currentfill}{rgb}{0.000000,0.000000,0.000000}%
\pgfsetfillcolor{currentfill}%
\pgfsetlinewidth{0.501875pt}%
\definecolor{currentstroke}{rgb}{0.000000,0.000000,0.000000}%
\pgfsetstrokecolor{currentstroke}%
\pgfsetdash{}{0pt}%
\pgfsys@defobject{currentmarker}{\pgfqpoint{0.000000in}{0.000000in}}{\pgfqpoint{0.020833in}{0.000000in}}{%
\pgfpathmoveto{\pgfqpoint{0.000000in}{0.000000in}}%
\pgfpathlineto{\pgfqpoint{0.020833in}{0.000000in}}%
\pgfusepath{stroke,fill}%
}%
\begin{pgfscope}%
\pgfsys@transformshift{0.773727in}{2.807684in}%
\pgfsys@useobject{currentmarker}{}%
\end{pgfscope}%
\end{pgfscope}%
\begin{pgfscope}%
\pgfsetbuttcap%
\pgfsetroundjoin%
\definecolor{currentfill}{rgb}{0.000000,0.000000,0.000000}%
\pgfsetfillcolor{currentfill}%
\pgfsetlinewidth{0.501875pt}%
\definecolor{currentstroke}{rgb}{0.000000,0.000000,0.000000}%
\pgfsetstrokecolor{currentstroke}%
\pgfsetdash{}{0pt}%
\pgfsys@defobject{currentmarker}{\pgfqpoint{-0.020833in}{0.000000in}}{\pgfqpoint{-0.000000in}{0.000000in}}{%
\pgfpathmoveto{\pgfqpoint{-0.000000in}{0.000000in}}%
\pgfpathlineto{\pgfqpoint{-0.020833in}{0.000000in}}%
\pgfusepath{stroke,fill}%
}%
\begin{pgfscope}%
\pgfsys@transformshift{4.077475in}{2.807684in}%
\pgfsys@useobject{currentmarker}{}%
\end{pgfscope}%
\end{pgfscope}%
\begin{pgfscope}%
\definecolor{textcolor}{rgb}{0.000000,0.000000,0.000000}%
\pgfsetstrokecolor{textcolor}%
\pgfsetfillcolor{textcolor}%
\pgftext[x=0.286111in, y=2.747684in, left, base]{\color{textcolor}{\rmfamily\fontsize{10.000000}{12.000000}\selectfont\catcode`\^=\active\def^{\ifmmode\sp\else\^{}\fi}\catcode`\%=\active\def
\end{pgfscope}%
\begin{pgfscope}%
\definecolor{textcolor}{rgb}{0.000000,0.000000,0.000000}%
\pgfsetstrokecolor{textcolor}%
\pgfsetfillcolor{textcolor}%
\pgftext[x=0.230556in,y=1.629739in,,bottom,rotate=90.000000]{\color{textcolor}{\rmfamily\fontsize{13.000000}{15.600000}\selectfont\catcode`\^=\active\def^{\ifmmode\sp\else\^{}\fi}\catcode`\%=\active\def
\end{pgfscope}%
\begin{pgfscope}%
\pgfpathrectangle{\pgfqpoint{0.773727in}{0.279167in}}{\pgfqpoint{3.303749in}{2.701145in}}%
\pgfusepath{clip}%
\pgfsetbuttcap%
\pgfsetmiterjoin%
\definecolor{currentfill}{rgb}{0.047059,0.364706,0.647059}%
\pgfsetfillcolor{currentfill}%
\pgfsetlinewidth{0.000000pt}%
\definecolor{currentstroke}{rgb}{0.000000,0.000000,0.000000}%
\pgfsetstrokecolor{currentstroke}%
\pgfsetstrokeopacity{0.000000}%
\pgfsetdash{}{0pt}%
\pgfpathmoveto{\pgfqpoint{1.660582in}{-226.701573in}}%
\pgfpathlineto{\pgfqpoint{1.660582in}{2.384956in}}%
\pgfpathlineto{\pgfqpoint{0.923897in}{2.384956in}}%
\pgfpathlineto{\pgfqpoint{0.923897in}{-226.701573in}}%
\pgfusepath{fill}%
\end{pgfscope}%
\begin{pgfscope}%
\pgfpathrectangle{\pgfqpoint{0.773727in}{0.279167in}}{\pgfqpoint{3.303749in}{2.701145in}}%
\pgfusepath{clip}%
\pgfsetbuttcap%
\pgfsetmiterjoin%
\definecolor{currentfill}{rgb}{0.000000,0.725490,0.270588}%
\pgfsetfillcolor{currentfill}%
\pgfsetlinewidth{0.000000pt}%
\definecolor{currentstroke}{rgb}{0.000000,0.000000,0.000000}%
\pgfsetstrokecolor{currentstroke}%
\pgfsetstrokeopacity{0.000000}%
\pgfsetdash{}{0pt}%
\pgfpathmoveto{\pgfqpoint{2.793943in}{-226.701573in}}%
\pgfpathlineto{\pgfqpoint{2.793943in}{0.379442in}}%
\pgfpathlineto{\pgfqpoint{2.057258in}{0.379442in}}%
\pgfpathlineto{\pgfqpoint{2.057258in}{-226.701573in}}%
\pgfusepath{fill}%
\end{pgfscope}%
\begin{pgfscope}%
\pgfpathrectangle{\pgfqpoint{0.773727in}{0.279167in}}{\pgfqpoint{3.303749in}{2.701145in}}%
\pgfusepath{clip}%
\pgfsetbuttcap%
\pgfsetmiterjoin%
\definecolor{currentfill}{rgb}{1.000000,0.584314,0.000000}%
\pgfsetfillcolor{currentfill}%
\pgfsetlinewidth{0.000000pt}%
\definecolor{currentstroke}{rgb}{0.000000,0.000000,0.000000}%
\pgfsetstrokecolor{currentstroke}%
\pgfsetstrokeopacity{0.000000}%
\pgfsetdash{}{0pt}%
\pgfpathmoveto{\pgfqpoint{3.927305in}{-226.701573in}}%
\pgfpathlineto{\pgfqpoint{3.927305in}{0.486552in}}%
\pgfpathlineto{\pgfqpoint{3.190620in}{0.486552in}}%
\pgfpathlineto{\pgfqpoint{3.190620in}{-226.701573in}}%
\pgfusepath{fill}%
\end{pgfscope}%
\begin{pgfscope}%
\pgfsetrectcap%
\pgfsetmiterjoin%
\pgfsetlinewidth{0.501875pt}%
\definecolor{currentstroke}{rgb}{0.000000,0.000000,0.000000}%
\pgfsetstrokecolor{currentstroke}%
\pgfsetdash{}{0pt}%
\pgfpathmoveto{\pgfqpoint{0.773727in}{0.279167in}}%
\pgfpathlineto{\pgfqpoint{0.773727in}{2.980312in}}%
\pgfusepath{stroke}%
\end{pgfscope}%
\begin{pgfscope}%
\pgfsetrectcap%
\pgfsetmiterjoin%
\pgfsetlinewidth{0.501875pt}%
\definecolor{currentstroke}{rgb}{0.000000,0.000000,0.000000}%
\pgfsetstrokecolor{currentstroke}%
\pgfsetdash{}{0pt}%
\pgfpathmoveto{\pgfqpoint{4.077475in}{0.279167in}}%
\pgfpathlineto{\pgfqpoint{4.077475in}{2.980312in}}%
\pgfusepath{stroke}%
\end{pgfscope}%
\begin{pgfscope}%
\pgfsetrectcap%
\pgfsetmiterjoin%
\pgfsetlinewidth{0.501875pt}%
\definecolor{currentstroke}{rgb}{0.000000,0.000000,0.000000}%
\pgfsetstrokecolor{currentstroke}%
\pgfsetdash{}{0pt}%
\pgfpathmoveto{\pgfqpoint{0.773727in}{0.279167in}}%
\pgfpathlineto{\pgfqpoint{4.077475in}{0.279167in}}%
\pgfusepath{stroke}%
\end{pgfscope}%
\begin{pgfscope}%
\pgfsetrectcap%
\pgfsetmiterjoin%
\pgfsetlinewidth{0.501875pt}%
\definecolor{currentstroke}{rgb}{0.000000,0.000000,0.000000}%
\pgfsetstrokecolor{currentstroke}%
\pgfsetdash{}{0pt}%
\pgfpathmoveto{\pgfqpoint{0.773727in}{2.980312in}}%
\pgfpathlineto{\pgfqpoint{4.077475in}{2.980312in}}%
\pgfusepath{stroke}%
\end{pgfscope}%
\begin{pgfscope}%
\definecolor{textcolor}{rgb}{0.000000,0.000000,0.000000}%
\pgfsetstrokecolor{textcolor}%
\pgfsetfillcolor{textcolor}%
\pgftext[x=1.292239in,y=2.426623in,,bottom]{\color{textcolor}{\rmfamily\fontsize{13.000000}{15.600000}\selectfont\catcode`\^=\active\def^{\ifmmode\sp\else\^{}\fi}\catcode`\%=\active\def
\end{pgfscope}%
\begin{pgfscope}%
\definecolor{textcolor}{rgb}{0.000000,0.000000,0.000000}%
\pgfsetstrokecolor{textcolor}%
\pgfsetfillcolor{textcolor}%
\pgftext[x=2.425601in,y=0.421109in,,bottom]{\color{textcolor}{\rmfamily\fontsize{13.000000}{15.600000}\selectfont\catcode`\^=\active\def^{\ifmmode\sp\else\^{}\fi}\catcode`\%=\active\def
\end{pgfscope}%
\begin{pgfscope}%
\definecolor{textcolor}{rgb}{0.000000,0.000000,0.000000}%
\pgfsetstrokecolor{textcolor}%
\pgfsetfillcolor{textcolor}%
\pgftext[x=3.558962in,y=0.528219in,,bottom]{\color{textcolor}{\rmfamily\fontsize{13.000000}{15.600000}\selectfont\catcode`\^=\active\def^{\ifmmode\sp\else\^{}\fi}\catcode`\%=\active\def
\end{pgfscope}%
\end{pgfpicture}%
\makeatother%
\endgroup%